\documentclass[aps,prb,reprint,showpacs,longbibliography,floatfix]{revtex4-2}

\usepackage{mathtools,amssymb,graphicx,units,bm}
\usepackage{amsmath}
\usepackage[plainpages=false,pdfpagelabels,colorlinks=true,linkcolor=red,urlcolor=blue,citecolor=blue,pdftitle={Title},pdfauthor={},pdfdisplaydoctitle=true,pdfduplex=DuplexFlipLongEdge]{hyperref}
\usepackage{soul} % to use \st for strikeout 
\usepackage[dvipsnames,usenames]{color}
\usepackage[normalem]{ulem}
\usepackage{color}
\usepackage{bbold} % to generate identity 1
\usepackage{float}
\emergencystretch=\maxdimen
\usepackage{dcolumn}% Align table columns on decimal point
\usepackage{bm}
\usepackage{physics}%
\usepackage{xcolor}
\usepackage{bbold} 
\usepackage{float}
\emergencystretch=\maxdimen
\def\s{\sigma}
\def\ave#1{\langle #1\rangle}
\newcommand{\avs}{\ave{\text{s}}}
\newcommand{\iv}{\mathbf{i}}
\newcommand{\jv}{\mathbf{j}}

\begin{document}

\normalem

\title{The half-filled extended Hubbard model on a square lattice:\\ Phase boundaries from determinant quantum Monte Carlo simulations
}
\author{Sebasti\~ao dos Anjos \surname{Sousa-J\'unior}}
%\email{sebastiaojr@pos.if.ufrj.br}
\author{Natanael C. Costa}
\author{Raimundo R. \surname{dos Santos}} 
\affiliation{Instituto de F\'\i sica, Universidade Federal do Rio de Janeiro
Cx.P. 68.528, 21941-972 Rio de Janeiro RJ, Brazil}

%%%%%%%%%%%%%%%%%%%%%%%%%%%%%%%%%%%%%%%%%%%%%%%%%%%%%%%%%%%%%%%%
\begin{abstract}
The extended Hubbard model (EHM) describes fermions on a lattice coupled through on-site, $U$, and first-neighbor, $V$, interactions. 
In the context of high-$T_c$ cuprates, antiferromagnetic fluctuations may lead to an attractive channel, hence to superconductivity. 
Despite interest in the two-dimensional version of the model, the current knowledge about the phase diagram is still far from complete. 
Here, we report on the results of extensive determinant quantum Monte Carlo simulations for this model at half filling, in which we have used the average sign of the product of fermionic determinants as an additional observable to locate critical points. 
We arrive at a ground state phase diagram in the $U$-$V$ plane in which the boundaries involving antiferromagnetic, charge-ordered, $s$- and $d$-wave superconductivity, and phase-separated phases are quantitatively set with good accuracy. 
We have also proposed a partial phase diagram, $T_c(U,V)$, featuring critical temperatures for the CDW and $s$-wave superconducting phases.
\end{abstract}
%%%%%%%%%%%%%%%%%%%%%%%%%%%%%%%%%%%%%%%%%%%%%%%%%%%%%%%%%%%%%%%%
\date{\today\ -- Version 3.1}
\maketitle

\section{Introduction}

Soon after the discovery of high-temperature cuprate superconductors \cite{Bednorz1986,Varma2020,Zhou2021}, a widespread consensus was formed around the idea that the basic physical mechanism leading to superconductivity was contained in the two-dimensional Hubbard model: pairing would emerge from strong antiferromagnetic (AFM) fluctuations arising from the competition between itinerancy (hopping) and localisation, the latter driven by an on-site repulsion of strength $U$ \cite{Anderson87,Emery87}. 
Despite its simplicity, the repulsive Hubbard model has eluded  an unambiguous characterisation of superconductivity. 
Subsequently, it was proposed \cite{Scalapino12} that antiferromagnetic fluctuations could actually lead to an additional effective nearest-neighbor attractive interaction, $V$. 
The ground state phase diagram for the half-filled extended Hubbard model (EHM) in one dimension has been known for some time; see, e.g.\,Ref.\,\cite{Lin1995}. 
For different values and signs of $U$ and $V$, one finds phases such as charge-density wave (CDW), spin-density wave (SDW), and $s$-wave superconductivity. 
A phase separated (PS) state appears for sufficiently large $-V$, e.g., half of the lattice with doubly occupied sites, and the other half empty, in the strong $-V$ regime.
Less conventional phases such as bond-ordered wave (BOW) \cite{Lin1995} and $p$-wave superconducting \cite{Xiao2022} have also been proposed to fit into the diagram.

In two dimensions, while one expects the ground state phase diagram at half filling to share similarities with that for the one-dimensional model, 
the picture is far from settled. 
Indeed, different methods agree with the existence of a CDW-AFM transition near $V=U/4$ \cite{zhang89,Huang2013,Vandelli20}. Nonetheless, the square lattice topology in principle allows for a wider range of pairing symmetries to be stabilized in the $V<0$ region, which has proved very hard to probe theoretically.
Weak-coupling methods \cite{Huang2013,Wolf18} may not fully capture the competition between the tendencies of phase separating and forming pairs of different symmetries, such as $s$ and $d$, also supported by dynamic cluster methods~\cite{Jiang2018,kundu2023}. 
Numerically exact diagonalisation of the Hamiltonian is currently restricted to $4\times4$ systems, so that the prediction of a $p$-wave paired state \cite{chen2022} may not stand for larger systems. 
Quantum Monte Carlo (QMC) simulations \cite{Blankenbecler81,hirsch85}, on the other hand, suffer from the infamous `minus-sign problem'  \cite{hirsch85,scalettar89,Kawashima2002,rrds2003,sorella2017}: 
 when the effective Boltzmann factor (given by a product of fermionic determinants) becomes negative, averages are taken  with it in absolute value, at the expense of dividing by its own average, $\ave{\text{s}}$, thus introducing excessive noise when $\ave{\text{s}}\ll 1$.
Recent studies of the EHM through finite temperature \cite{Sushchyev2022} and projective QMC simulations~\cite{Yao22} were carried out using  complex Hubbard-Stratonovich fields (CHSF; see Appendices \ref{App:RHSF} and \ref{App:CHSF}), so that calculations were restricted to the regime $|V|\leq |U|/4$, where $\ave{s}=1$.
As a consequence, since issues such as a precise location of the CDW-AFM phase boundary, the transition between superconducting phases with different pairing states, and the complete boundary to the phase-separated regime all lie outside this range, they have not yet been thoroghly probed with QMC simulations.
The inescapable conclusion is that so far the overall knowledge of the ground state EHM phase diagram is, at best, semi-qualitative. 
One certainly needs to grasp the half-filled regime before comparing with the doped case, closer to the cuprate superconductors.

%%%%%%%%%%%%%%%%%%%%%%%%%%%%%%%%%%%%%%%%%%%%%%% Fig 1 %%%%%%%%%%%%%%%
\begin{figure*}[t]
    \centering
    \includegraphics[scale=0.5]{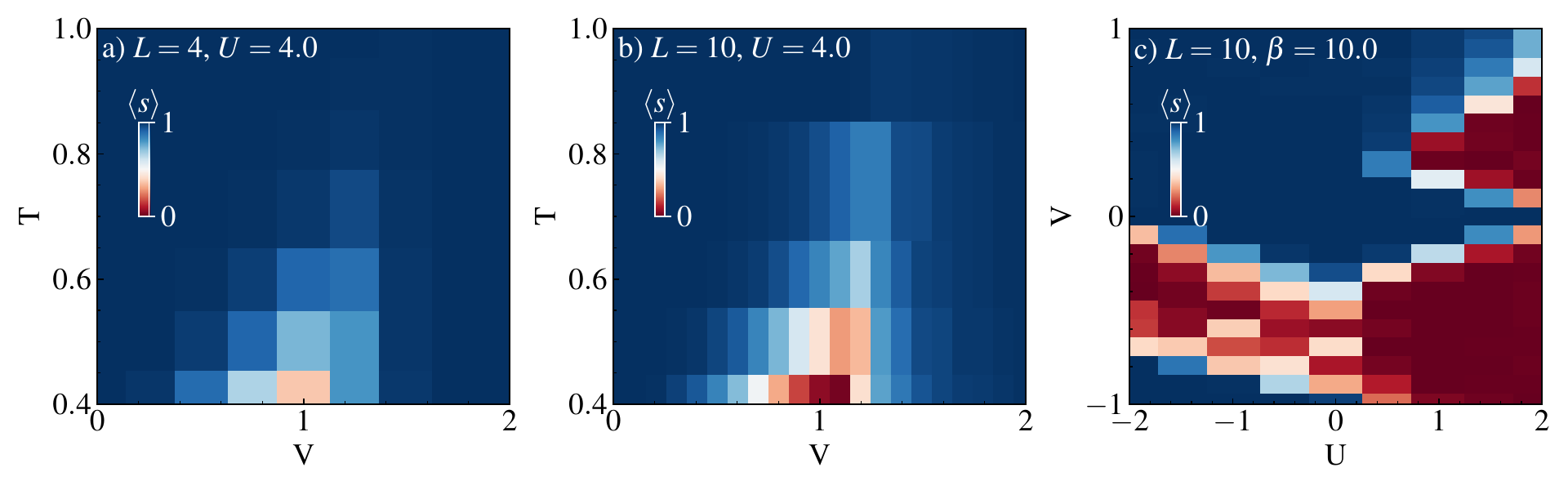}
    \caption{Contour plots for the average sign of the EHM at half filling on $L\times L$ lattices: (a) as a function of $T$ and $V$, for $L=4$ and $U=4t$; (b) same as (a), but for $L=10$; (c) as a function of $U$ and $V$, for $L=10$ and fixed inverse temperature,   $\beta=10$.}
    \label{fig:sign}
\end{figure*}
%%%%%%%%%%%%%%%%%%%%%%%%%%%%%%%%%%%%%%%%%%%%%

In addition to ground state properties, there are issues related to phase transitions at finite temperatures which are worth examining. 
Indeed, the addition of a finite nearest-neighbor interaction to the attractive Hubbard model breaks the degeneracy of the superconducting (SUC) and CDW phases at half filling. 
This bypasses the Mermin–Wagner–Hohenberg theorem \cite{Mermin1966,Hohenberg1967} which rules out their simultaneous ordering at finite temperatures.
One may therefore expect an increase in $T_c$ for both CDW ordering and superconductivity, respectively for $V>0$ and $V<0$; a feature worth probing is how steep is this increase.

With the purpose of shedding light into these unresolved issues, here we perform a detailed investigation of the EHM at half filling through determinant QMC (DQMC) simulations. 
From the outset we stress that the above mentioned minus sign problem is mitigated through two procedures. 
One is the use of CHSF in the range $|V|\leq |U|/4$ to obtain several response functions whose behavior would be otherwise unattainable, while the other resorts to the recent observations \cite{Wessel2017,Mondaini2022,Mondaini2023} that a severely degraded $\ave{\text{s}}$ may actually be used to pinpoint ground state phase transition points and boundaries.
The layout of the paper is as follows.
In Sec.\,\ref{sec:modelmethods}, we present the Hamiltonian and highlight the DQMC method (additional details are left to the Appendix).
Section\,\ref{sec:GS} presents the results for the ground-state transitions,
while Sec.\,\ref{sec:critical_temps} focuses on finite temperature transitions. 
Finally, Sec.\,\ref{sec:conclusions} summarizes our findings.

%%%%%%%%%%%%%%%%%%%%%%%% II. Model and Methods %%%%%%%%%%%%%%%%%%%%%%%% 

\section{Model and Methods}
\label{sec:modelmethods}

The Hamiltonian for the EHM reads,  
\begin{align}
    \mathcal{H}   = & -t \sum_{\langle \textbf{i,j} \rangle} (c_{\textbf{i}\sigma}^\dagger c^{\phantom{\dagger}}_{\textbf{j}\sigma} + \text{H.c.})  - \mu \sum_{\textbf{i},\sigma} n_{\textbf{i},\sigma}\nonumber\\
    & + U\sum_\textbf{i} n_{\textbf{i}\uparrow}n_{\textbf{i}\downarrow} +V\sum_{\langle \textbf{i,j} \rangle} n_{\textbf{i}}n_{\textbf{j}}, 
    \label{eq:EHM}
\end{align}
where $\iv$ and $\jv$ denote sites of a square lattice, with $\ave{\iv,\jv}$ restricting the sums to nearest neighbor (NN) sites. 
In standard second-quantized notation, the first term describes fermionic hopping (energy scale $t$),  the second controls the band filling through the chemical potential, $\mu$, while the third and fourth terms describe the on-site and NN interactions, with strengths $U$ and $V$, respectively. 
Hereafter, the chemical potential is set to $\mu=U/2+4V$ to yield a half-filled band (by virtue of particle-hole symmetry), and energies are expressed in units of $t$.     

In the DQMC method \cite{hirsch85,scalettar89,Kawashima2002,rrds2003,sorella2017}, discrete Hubbard-Stratonovich (HS) transformations \cite{Hirsch83} are employed to express the quartic interactions in quadratic forms. 
This leads to the introduction of auxiliary fields \cite{zhang89,golor2015,Yao22}; see Appendices \ref{App:RHSF} and \ref{App:CHSF}. 
The non-commutation between the one-body and the two-body terms of the Hamiltonian is taken care of through a Suzuki-Trotter decomposition, which adds an imaginary-time dimension, $L_\tau=\beta/\Delta\tau$, with $\beta$ being the inverse temperature, and $\Delta\tau$ the discrete time step. The trace over the fermionic degrees of freedom can then be performed, leading to the product of determinats alluded to before, which weigh the configurations of the HS Ising fields by importance sampling, as in usual Monte Carlo methods \cite{hirsch85,scalettar89,Kawashima2002,rrds2003,sorella2017}.

The type of ordering is characterised by quantities such as the (staggered) charge structure factor,
\begin{align}
    S_{\text{cdw}} = \frac{1}{N} \sum_{\textbf{i}, \textbf{j}} (-1)^{|\iv-\jv|} \langle(n_{\textbf{i},\uparrow} + n_{\textbf{i},\downarrow} )(n_{\textbf{j},\uparrow} + n_{\textbf{j},\downarrow} ) \rangle,
\label{eq:scdw}
\end{align}
the antiferromagnetic structure factor,
\begin{align}
S_{\text{afm}}  = \frac{1}{N} \sum_{\textbf{i}, \textbf{j}} 
(-1)^{|\iv-\jv|}
\langle S^{z}_{\mathbf{i}} S^{z}_{\mathbf{j}} \rangle,
\label{eq:ss}
\end{align}
and the pairing structure factor,
\begin{align}
P_{\text{sc}}(\alpha) = \frac{1}{N} \sum_{\textbf{i,j}}  \langle \Delta_\alpha (\textbf{i}) \Delta^\dagger_\alpha (\textbf{j}) \rangle,
\label{eq:pairing_cor}
\end{align}
with 
\begin{equation}
	\Delta _\alpha (\textbf{i}) = \sum_\textbf{a} f_\alpha (\textbf{a}) c_{\textbf{i}\downarrow} c_{\textbf{i+a}\uparrow}. 
\end{equation}
In the above equations, $N = L \times L$ is the number of sites for a linear size $L$, $(-1)^{|\iv-\jv|}$ is $\pm 1$ if $\iv$ and $\jv$ are on the same or opposite sublattices, $S^{z}_{\mathbf{i}} = (n_{\mathbf{i} \uparrow} - n_{\mathbf{i} \downarrow})$ is the $z$-component of the spin operator, and $f_\alpha (\textbf{a})$ is the form factor for a given pair-wave symmetry, $\alpha={s,d,p}$~\cite{white89}.
%\color{blue}
In some circumstances, it is more appropriate to calculate the pair susceptibility
\begin{align}
\chi^{\alpha}_{\rm sc}(\beta) = \frac{1}{N} \sum_{\textbf{i,j}} 
\int^{\beta}_{0} \mathrm{d}\tau \,
\langle \Delta_{\mathbf{i},\alpha}^{\phantom{\dagger}}(\tau)\Delta^{\dagger}_{\mathbf{j},\alpha}(0) \rangle~,
\end{align}
with
$ \Delta_{\alpha}(\mathbf{i},\tau) = \sum_{\mathbf{a}} f_{\alpha}(\mathbf{a})
c^{\phantom{\dagger}}_{\mathbf{i}\downarrow}(\tau) 
c^{\phantom{\dagger}}_{\mathbf{i}+\mathbf{a} \uparrow}(\tau)$, and
$c_{\mathbf{i}\sigma}(\tau)=e^{\tau\cal{H}}c_{\mathbf{i}\sigma}e^{-\tau\cal{H}}$,
which provides a stronger signal of pairing properties.
As mentioned before, $\avs$, which is automatically calculated in the simulations, will also play a crucial role in our analyses.

When $V\neq 0$ one cannot guarantee that $\avs=1$ even at half filling.
As shown in Figs.\,\ref{fig:sign}(a) and (b), the average sign decreases as the temperature is lowered, and worsens as $L$ increases.
Nonetheless, for both system sizes, $\avs$ approaches zero faster near $V = 1$. 
Recalling that the phase boundary for the SDW-CDW transition is $V_c\approx U/4$, the minimum of $\avs$ at $V = 1$ can hardly be regarded as fortuitous. 
A mapping of $\ave{\text{s}}$ in the \linebreak $V$-$U$ plane at a fixed low temperature is shown in Fig.\,\ref{fig:sign}\,(c).
Apart from the second quadrant ($U<0$,$V>0$), one sees that there are regions with $\ave{\text{s}}\ll1$ surrounded by less severe ones. 
In what follows we combine analyses of $\avs$ with the quantities defined by Eqs.\,\eqref{eq:scdw}-\eqref{eq:pairing_cor} to first obtain the ground state phase boundaries. 

%%%%%%%%%%%%%%%%%%%%%%%% III. Ground State Results %%%%%%%%%%%%%%%%%%%%%%%% 

\section{Ground State Results}
\label{sec:GS}
\subsection{AFM-CDW transition}
\label{ssec:AFM-CDW}

%%%%%%%%%%%%%%%%%%%%%%%%%%%%%%%%%%%%%%%%%%%%%%% Fig 2 %%%%%%%%%%%%%%%
\begin{figure}[t]
    \centering
    \includegraphics[scale=0.5]{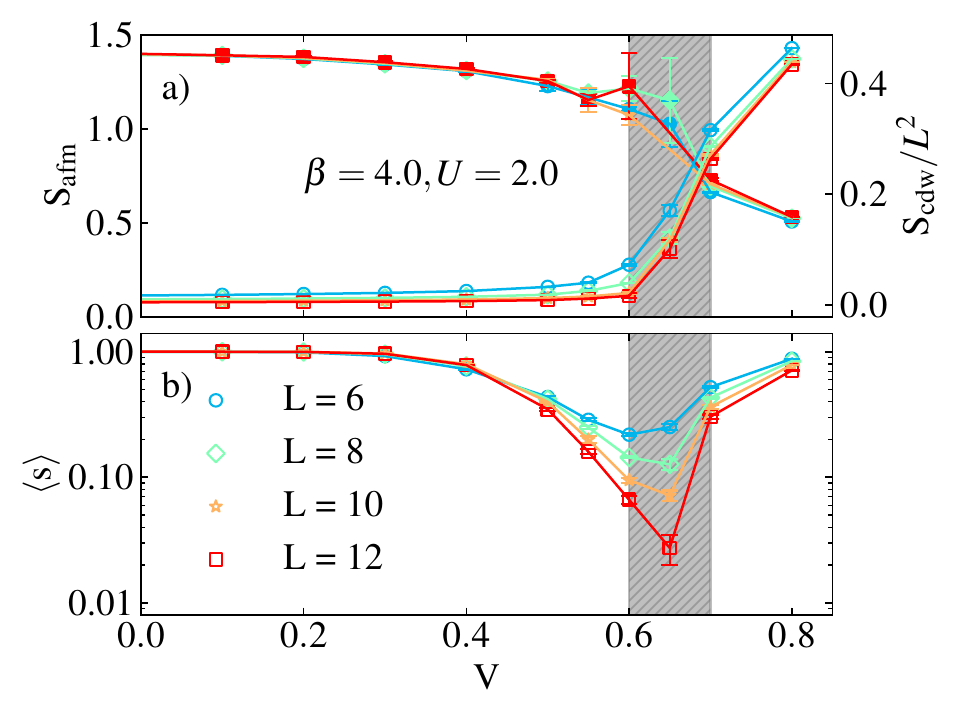}
    \caption{(a) Spin structure factor (filled symbols, left vertical scale) and scaled charge structure factor (empty symbols, right vertical scale), and (b) average sign as functions of $V$.
    All data are for fixed $U=2$ and $\beta=4$, and different linear system sizes, $L$.
    The grey bar highlights the region where $\avs$ dips.
    }
    \label{fig:avesB4}
\end{figure}
%%%%%%%%%%%%%%%%%%%%%%%%%%%%%%%%%%%%%%%%%%%%%%%%%%%%%%%%%%%%%%%%%%

We start with the transition between the AFM (Mott) phase and the CDW phase.
As indicated in Fig.\,\ref{fig:avesB4}\,(a), increasing $V$ with fixed $U$ causes a sharp decrease of $S_\text{afm}$ and a sharp increase of $S_\text{cdw}$ thus signalling a phase transition near $V=0.6$.
Figure \ref{fig:avesB4}\,(b) shows $\avs$ calculated with real HS fields to illustrate that the change in $S_\text{afm}$ and $S_\text{cdw}$ is accompanied by a sharp dip. 
Given the amount of data gathered so far relating a degraded $\avs$ with quantum critical points \cite{Wessel2017,Mondaini2022,Mondaini2023,Lima2023}, we estimate the critical point for for $U=2$ as $V_c=0.65\pm0.05$, where the error reflects the $V$ increments in Fig.\,\ref{fig:avesB4}\,(b).

\begin{figure}[t]
    \centering
    \includegraphics[scale=0.5]{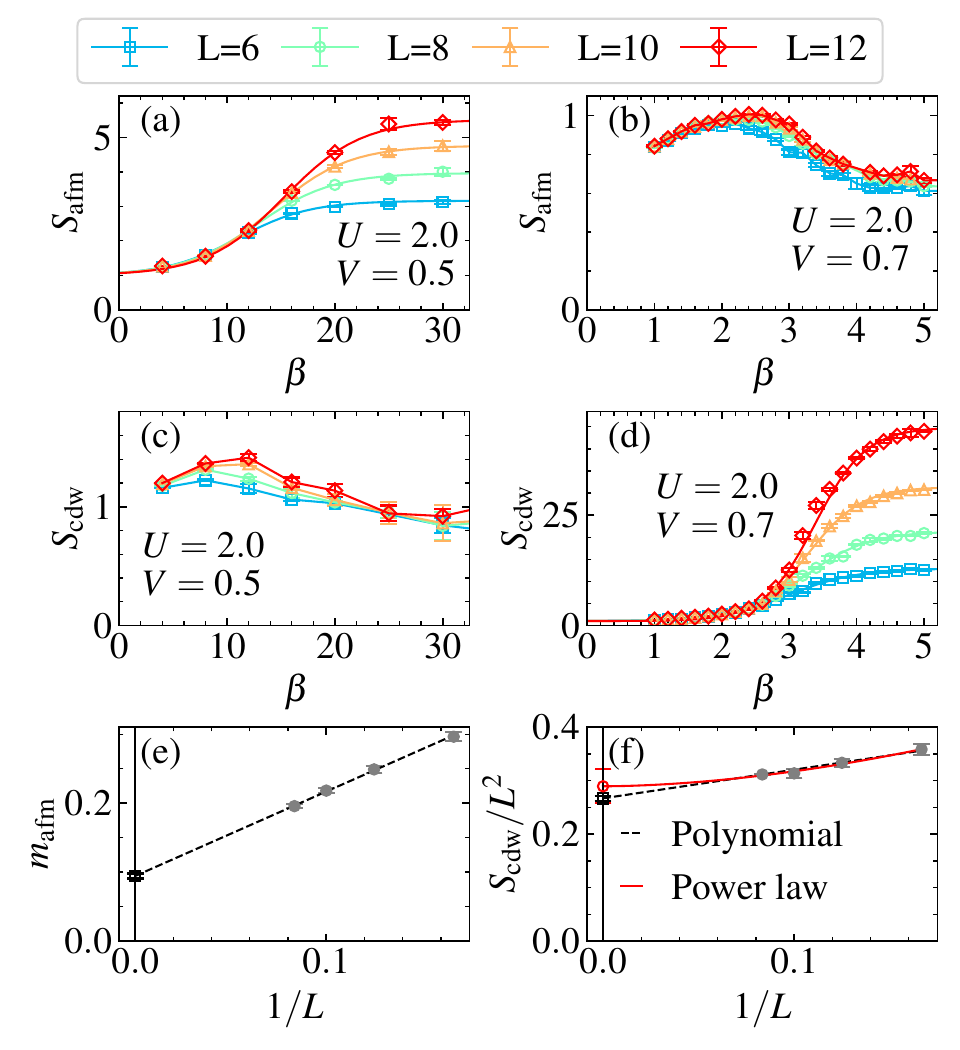}
    \caption{
    (a) Spin structure factor as a function of inverse temperature for $U=2$, different lattice sizes, and $V=0.5$. (b) Same as (a), but for $V=0.7$. (c) Charge structure factor as a function of inverse temperature for $U=2$, different lattice sizes, and $V=0.5$. (d) Same as (c) but for $V=0.7$. 
    (e) Finite-size analyses for the staggered magnetization (see text). 
    (f) Same as (e), but for the CDW structure factor. 
    }
    \label{fig:FSSU2}
\end{figure}

In addition, we recall that for $U=2$ the sign-free region for CHSF corresponds to $V\leq 0.5$, so that we may obtain these structure factors at very low temperatures for $V=0.5$; see Figs.\,\ref{fig:FSSU2}(a) and (c). 
The stabilized values (i.e. when $\beta\to\infty$) of $S_\text{afm}$ in Fig.\,\ref{fig:FSSU2}(a) can then be used in the scaling \textit{ansatz} \cite{Huse88}, $m_\text{afm} \sim \sqrt{S_\text{afm}/L^2}$, to extract the ground state staggered magnetization, $m_\text{afm}$, when $L\to\infty$, as in Fig.\,\ref{fig:FSSU2}(e).
By contrast, in the low temperature regime $S_\text{cdw}$ is practically independent of $L$, indicating the suppression of CDW correlations.
This analysis thus confirms that for $V=0.5$ the ground state is antiferromagnetic. 

%%%%%%%%%%%%%%%%%%%%%%%%%%%%%%%%%%%%%%%%%%%%%%% Fig 3 %%%%%%%%%%%%%%%
%%%%%%%%%%%%%%%%%%%%%%%%%%%%%%%%%%%%%%%%%%%%%%%%%%%%%%%%%%%%%%%%%%

%%%%%%%%%%%%%%%%%%%%%%%%%%%%%%%%%%%%%%%%%%%%%%% Fig 4 %%%%%%%%%%%%%%%
\begin{figure*}[t]
    \centering
    \includegraphics[scale=0.5]{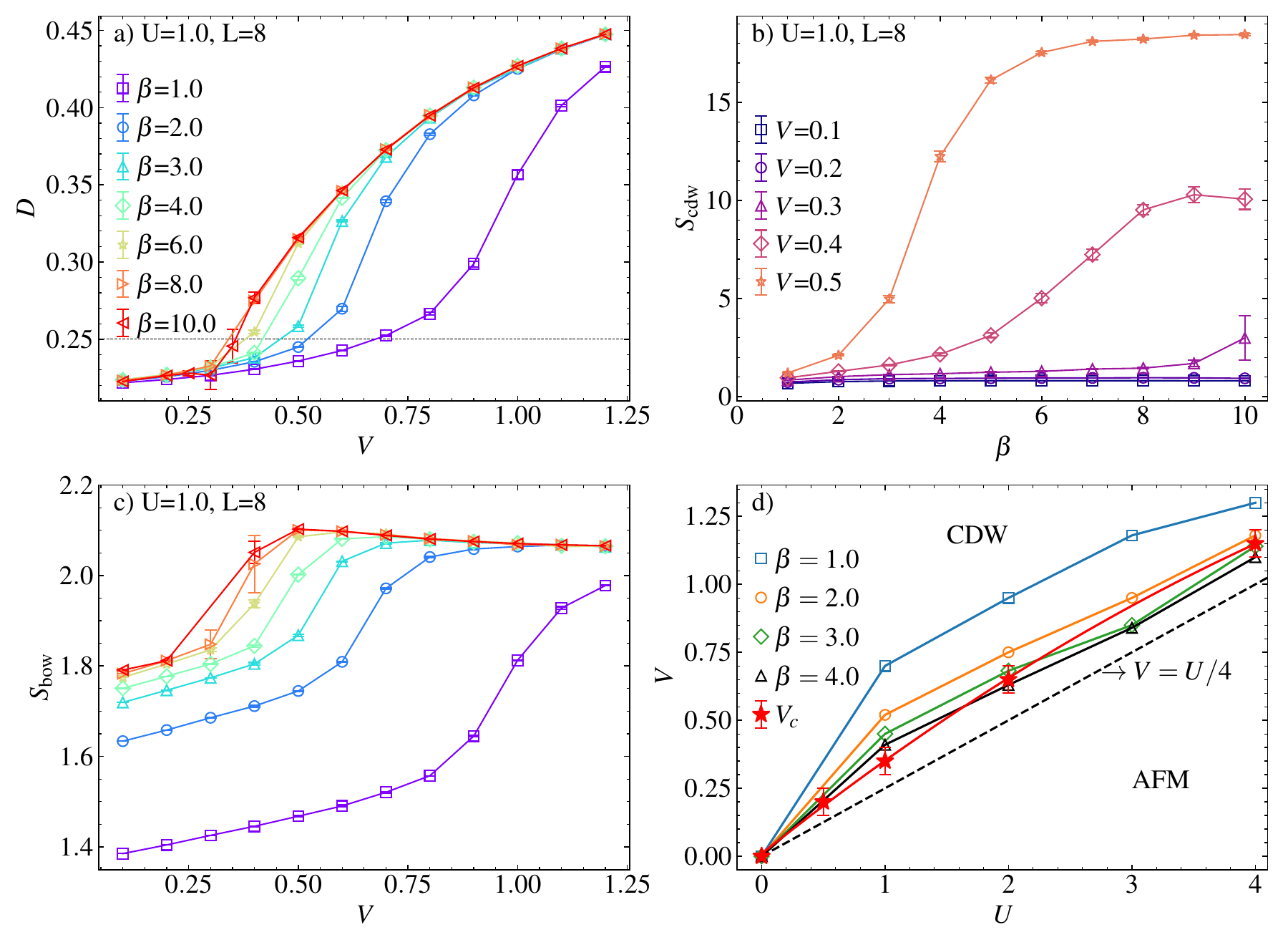}
    \caption{(a) Double occupancy as a function of $V$, for different values of $\beta$. 
    (b) Charge structure factor as a function of $\beta$, for different values of $V$. 
    (c) BOW structure factor as a function of $V$ for the same values of $\beta$ as in (a). 
    (d) First quadrant ground state phase diagram for the EHM: empty symbols indicate points where $D\sim 0.25$, for different values of $\beta$: filled stars indicate the critical points obtained from analyses similar to Fig.\,\ref{fig:avesB4}; the dashed line is the strong coupling critical line, $V=U/4$.
    }
    \label{fig:double_occ}
\end{figure*}
%%%%%%%%%%%%%%%%%%%%%%%%%%%%%%%%%%%%%%%%%%%%%

Let us now discuss the data for $V=0.7$, still with $U=2$. 
Although these parameters lie outside the range of applicability of the CHSF, Fig.\,\ref{fig:avesB4}\,(b) shows that $\avs\gtrsim0.3$, which allows us to calculate the correlation functions quite confidently; see Appendix \ref{App:sign}.
Figures \ref{fig:FSSU2}(b) and (d) show that the roles of $S_\text{afm}$ and $S_\text{cdw}$, as far as the $L$-dependence is concerned, have been inverted in relation to the case $V=0.5$: while the former hardly depends on $L$, the latter increases steadily with $L$. 
Accordingly, Fig.\,\ref{fig:FSSU2}(f) shows that $S_\text{cdw}/L^2$ extrapolates to a finite value as $L\to\infty$, signaling the onset of long-ranged CDW order.  
These analyses are in perfect agreement with our estimates based on Fig.\,\ref{fig:avesB4}.
For other values of $U$, we follow the same analyses, and it turned out that the behavior is quite similar, leading to $V_c=0.3\pm 0.1$ for $U=1$, and $V_c=1.1\pm 0.1$ for $U=4$.

Further checks can be carried out by examining the double occupancy, $D = \ave{n_\uparrow n_\downarrow}$ in the region of small $U$. 
Since an AFM state should yield $D\lesssim 0.25$, while a CDW state leads to $D\gtrsim 0.25$, close to a AFM-CDW transition one may expect $D\approx 0.25$, the noninteracting result. As a first-order phase transition may occur between an AFM and a CDW phase, the finite temperature behavior of $D$ would exhibit a sharp change around the transition region. Indeed, this was employed in literature to estimate the critical region of other models with AFM-CDW transition\,\cite{Nowadnick12,Johnston13,Costa20}.
Figure \ref{fig:double_occ}(a) shows $D$ as a function of $V$, for $U=1$, and different inverse temperatures. 
The curves for $\beta=8$ and $10$ are practically the same, and cross the $D=0.25$ horizontal line at $V=0.35\pm0.05$.
For comparison, in Fig.\,\ref{fig:double_occ}(b) we plot $S_\text{cdw}(\beta)$ for different values of $V$, and the charge correlations are only enhanced at low temperatures if $V \gtrsim 0.3\pm 0.1$, consistently with the estimates from $D$. 

It is also instructive to check whether a bond-ordered wave (BOW) phase can be identified near the AFM-CDW transition region, similarly to the one-dimensional case \cite{Lin1995,Ferreira22}.
To this end, we calculate the corresponding bond correlation function \cite{Xing2021}, defined as
\begin{align}
S_\text{bow} = \frac{1}{N} \sum_{\textbf{i}, \textbf{j},\alpha=x,y} (-1)^{|\iv-\jv|}
 \left\langle K_{\textbf{i,i}+\hat{\alpha}}^\sigma K_{\textbf{j,j}+\hat{\alpha}}^\sigma \right\rangle ~,  
\end{align}
where $K_{\textbf{i,j}}^\sigma \equiv (c_{\textbf{i}\sigma}^\dagger c_{\textbf{j}\sigma} + \text{H.c.}) $.  
In Figure \ref{fig:double_occ}(c) $S_\text{bow}$ is depicted as a function of $V$ for various values of $\beta$. 
Although the BOW correlations display a maximum at low temperatures, there is no unambiguous signature of enhancement, to the point of characterizing the sought BOW phase, especially because this maximum occurs deep in the CDW phase.

We conclude our discussion by summarizing the findings of the AFM-CDW transition in Fig.\,\ref{fig:double_occ}(d). 
Empty symbols represent points where $D\sim 0.25$ for various values of $\beta$.
As the temperature decreases, these curves approach the transition line passing through the solid symbols, as determined by correlation functions and $\avs$.
The key observation from Fig.\,\ref{fig:double_occ}(d) is that the transition curve lies slightly above the line $V = U/4$. 
This is in agreement with recent work using Cluster Dynamical Mean Field theory~\cite{kundu2023}, but in disagreement with an early mean-field approach \cite{Dagotto94}, which predicts $V_c = U/4$.
Unfortunately, we are unable to compare with data from previous QMC simulations \cite{Sushchyev2022,Yao22} since their data are restricted to $|V|\leq |U|/4$, thus excluding the AFM-CDW transition.

%%%%%%%%%%%%%%%%%%%%%%%%%%%%%%%%%%%%%%%%%%%%%%% Fig 5 %%%%%%%%%%%%%%%
\begin{figure}[t]
    \centering 
    \hskip -0.3cm
    \includegraphics[scale=0.5]{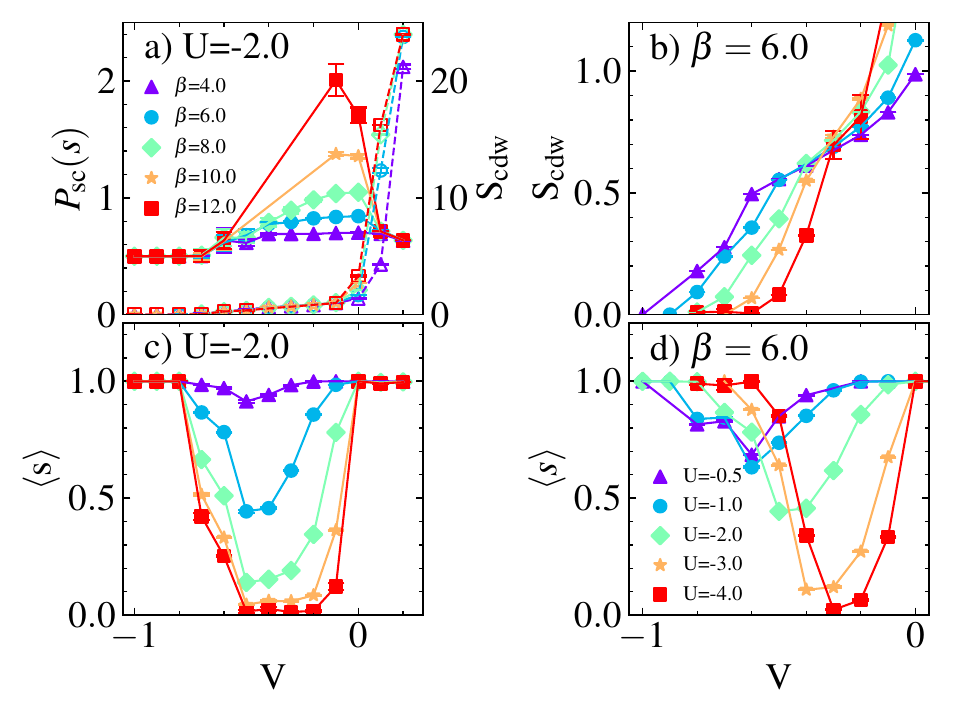}
    \caption{
    (a) $s$-wave pairing structure factor (full lines, left vertical axis) and charge structure factor (dashed line, right vertical axis), and (c) average sign as functions of $V$ at fixed $U$, for different temperatures; 
    (b) charge structure factor [notice the different scale from (a)] and (d) average sign as functions of $V$, at fixed inverse temperature, for different values of the on-site attraction. All data are for $L=8$.
    }
    \label{fig:NUL8}
\end{figure}
%%%%%%%%%%%%%%%%%%%%%%%%%%%%%%%%%%%%%%%%%%%%%

\subsection{CDW-SUC transition}

Let us now focus on the second ($U<0$, $V>0$) and third ($U<0$, $V<0$) quadrants of the parameter space.
Figure \ref{fig:NUL8}\,(a) shows that the charge structure factor decreases steadily as one crosses the $V=0$ line towards $V<0$, at fixed $U=-2$; by contrast, the $s$-wave pairing structure factor  increases steadily with increasing $\beta$ on the $V<0$ side.
This behavior is consistent with the fact that exactly at $V=0$ one reaches the attractive Hubbard model, which definitely displays a CDW  state coexisting with an $s$-wave superconducting state \cite{Micnas90, fontenele2022}: one deals effectively with a three-component order parameter, one for CDW and two for the superconductivity.
Further decrease in $V$ causes a suppression of $s$-wave pairing correlations, as one enters the PS region, at $V= -0.45\pm 0.05$ for $U=-2$.

Figure \ref{fig:NUL8}\,(c) shows $\avs$ calculated with real HS fields for us to follow its role as a phase transition marker; we recall that simulations with CHSF yield a constant $\avs=1$ between $V=-0.5$ and $V=0.5$ (not shown). 
Interestingly, Figure \ref{fig:NUL8}\,(c) shows that as the temperature decreases, the dip in $\avs$ deepens and widens throughout the superconducting phase; in the PS region, $\avs$ returns to 1.
Thus the sharp drops in $\avs$ indicate the boundaries of the superconducting phase with the CDW and PS regions.
Figure \ref{fig:NUL8}\,(b) provides an interesting insight into the behavior of the charge structure factor, by keeping the temperature fixed at $\beta=6$, and examining how the plots evolve with $U$. 
Two regimes can be distinguished: one in which charge correlations are enhanced as $|U|$ increases, and another in which they decrease as $|U|$ increases. 
The analysis of Fig.\,\ref{fig:NUL8}\,(a) hence suggests that this change indicates the entrance into the PS region.
This is again in accordance with the behavior of $\avs$ at fixed $\beta$ and for different values of $U$, depicted in Fig.\,\ref{fig:NUL8}\,(d): the dip occurs at smaller values of $V$ as $U$ increases.
These findings are summarised in Fig.\,\ref{fig:phasediag}.

%%%%%%%%%%%%%%%%%%%%%%%%%%%%%%%%%%%%%%%%%%%%%%% Fig 6 %%%%%%%%%%%%%%%
\begin{figure}[t]
    \centering    
    \includegraphics[scale=0.7]{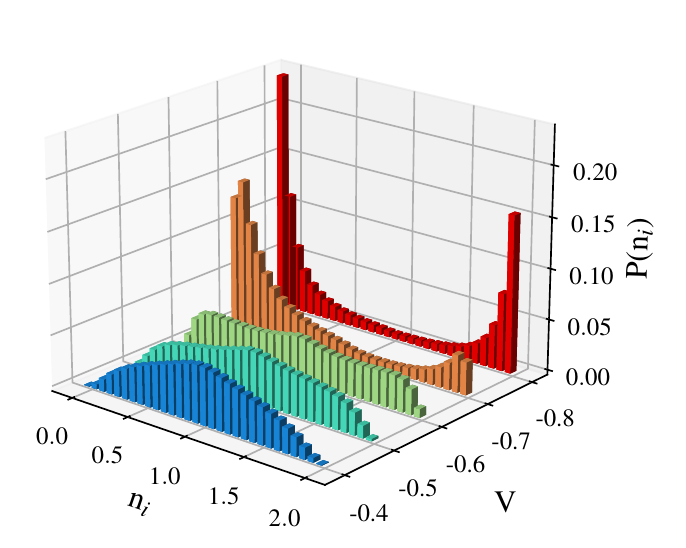}
    \caption{Density probability distribution for several values of $V$ with $U=-0.5$ and $\beta=6.0$, for $L=8$. As the intensity of the attractive extended interaction increases the shape of the histograms undergoes from a single peaked to a double peaked distribution.}
    \label{fig:histogram}
\end{figure}
%%%%%%%%%%%%%%%%%%%%%%%%%%%%%%%%%%%%%%%%%%%%%

The transition to the PS state deserves a complementary look, by examining the density distribution shown in Fig.\,\ref{fig:histogram}, generated by collecting the values of $n_\iv$ over the DQMC runs.  
For $V=0$, the distributions are represented by singly peaked histograms (not shown), centred at $n=1$. 
As $V$ decreases, the distributions first broaden, still with a peak at $n=1$, but a change to doubly-peaked at $n=0$ and $n=2$ takes place, interpreted as a signature of a phase separated state. 
In Fig.\,\ref{fig:histogram}, this occurs at $V\approx -0.6$, which also marks the dip in $\avs$ for $U=-0.5$; see Fig.\,\ref{fig:NUL8}\,(d).

%%%%%%%%%%%%%%%%%%%%%%%%%%%%%%%%%%%%%%%%%%%%%%% Fig 7 %%%%%%%%%%%%%%%
\begin{figure}[t]
    \centering
    \includegraphics[scale=0.5]{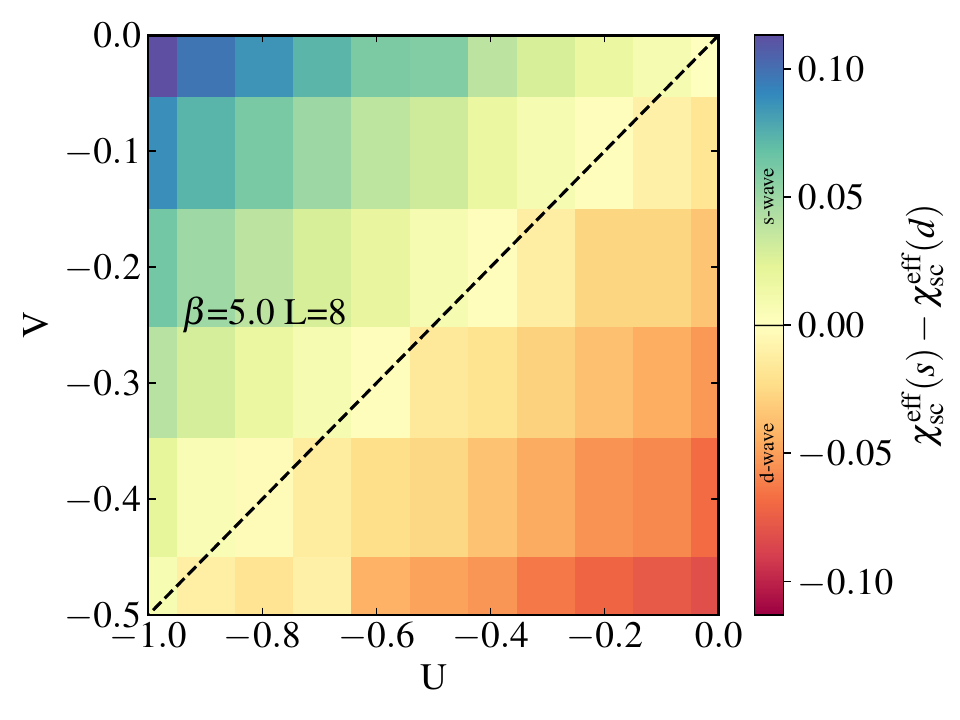}
    \caption{a) Contour plot of the difference between $s$-wave and $d$-wave effective pairing susceptibilities for $\beta=5.0$ and $L=8$.}
    \label{fig:peff}
\end{figure}
%%%%%%%%%%%%%%%%%%%%%%%%%%%%%%%%%%%%%%%%%%%%%

\subsection{Superconducting pairing symmetries}
Returning to the SC state, we must settle the issue of pairing symmetries. 
To this end, we start with the difference between the effective pairing susceptibilities with $s$ and $d$ symmetries, which presumably dominate the third quadrant.
At this point, we investigate the effective pairing (vertex) susceptibility, i.e., $\chi^{\rm eff}_{\rm sc} (\alpha) = \chi_{\rm sc} (\alpha) - \bar{\chi}_{\rm sc} (\alpha)$, with $\bar{\chi}_\text{sc}(\alpha)$ being the noninteracting susceptibility~\cite{white89}.
A positive (negative) response of $\chi^{\rm eff}_{\rm sc} (\alpha)$ signals an enhancement (weakening) of pairing correlations for the corresponding $\alpha$-wave symmetry.
From Fig.\,\ref{fig:peff} we see that the region above the line $V = U/2$ is dominated by $s$-wave pairing, while the region below the line is dominated by $d$-wave pairing.
Interestingly, $\avs$ shows no dip at this transition; we may attribute this to the lack of change in the number of components of the order parameter on either side of the transition.

\begin{figure}[t]
    \hskip -0.3cm
    \centering
    \includegraphics[scale=0.55]{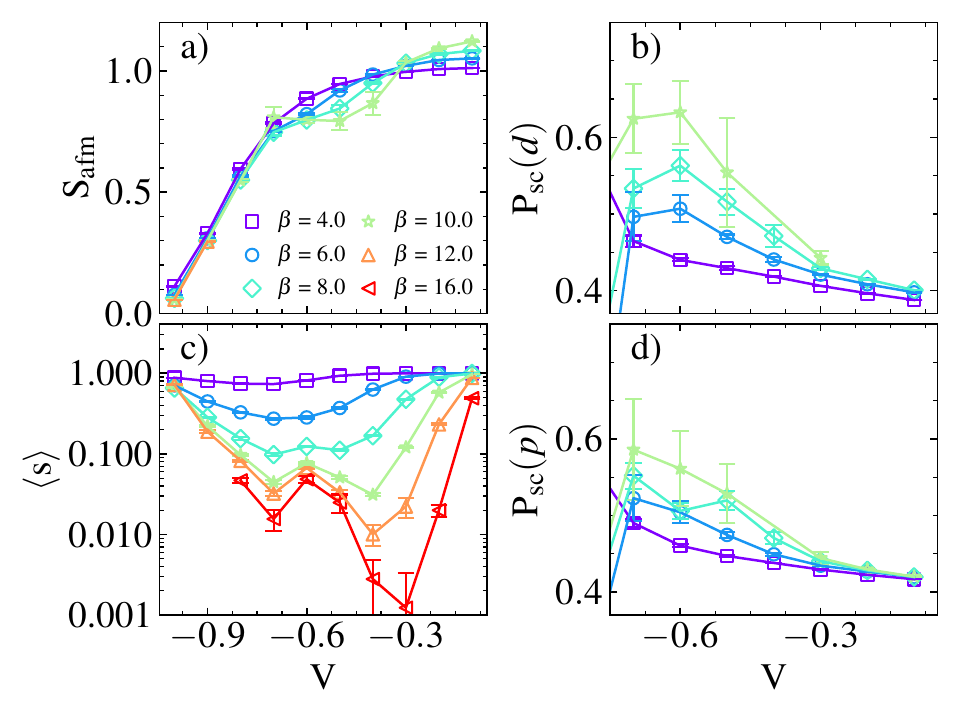}
    \caption{a) Spin structure factor,  b) $d$-wave pairing correlation function  c) Average sign and d) $p$-wave pairing correlation function as a function of $V$ for $U=0.8$ with $L=6$. 
    }
    \label{fig:avesL6}
\end{figure}

Moving on to the fourth quadrant, $V<0$ and $U>0$, Fig.\,\ref{fig:avesL6}(a) follows how the spin structure factor changes as $V$ is increased. 
Below $V\approx -0.7$, $S_\text{afm}$ is quite insensitive to the temperature. 
For $-0.7\lesssim V \lesssim -0.25$, $S_\text{afm}$ actually decreases as the temperature decreases, while the superconducting structure factors are enhanced in this interval [see Figs.\,\ref{fig:avesL6}(b) and (d)], with $P_\text{sc}(d)$ tending to dominate over $P_\text{sc}(p)$.
The dip in $\avs$ at $V\approx -0.7$ provides additional support to the interpretation of a PS--$d$-wave transition at this point in the diagram.
Beyond $V\approx -0.25$, $S_\text{afm}$ increases as the temperature is lowered, which is also accompanied by a dip in $\avs$: this signals a $d$-wave--AFM transition. 
By repeating these analyses for other values of $U>0$, we obtain the critical curve in Fig.\,\ref{fig:phasediag}, in which the error stems from the resolution of the crossings in Fig.\,\ref{fig:avesL6}(a).

%%%%%%%%%%%%%%%%%%%%%%%%%%%%%%% Fig 11 %%%%%%%%%%%%%%%
%%%%%%%%%%%%%%%%%%%%%%%%%%%%%%%%%%%%%%%%%%%%%%

\begin{figure}[t]
    \centering
    \includegraphics[scale=0.5]{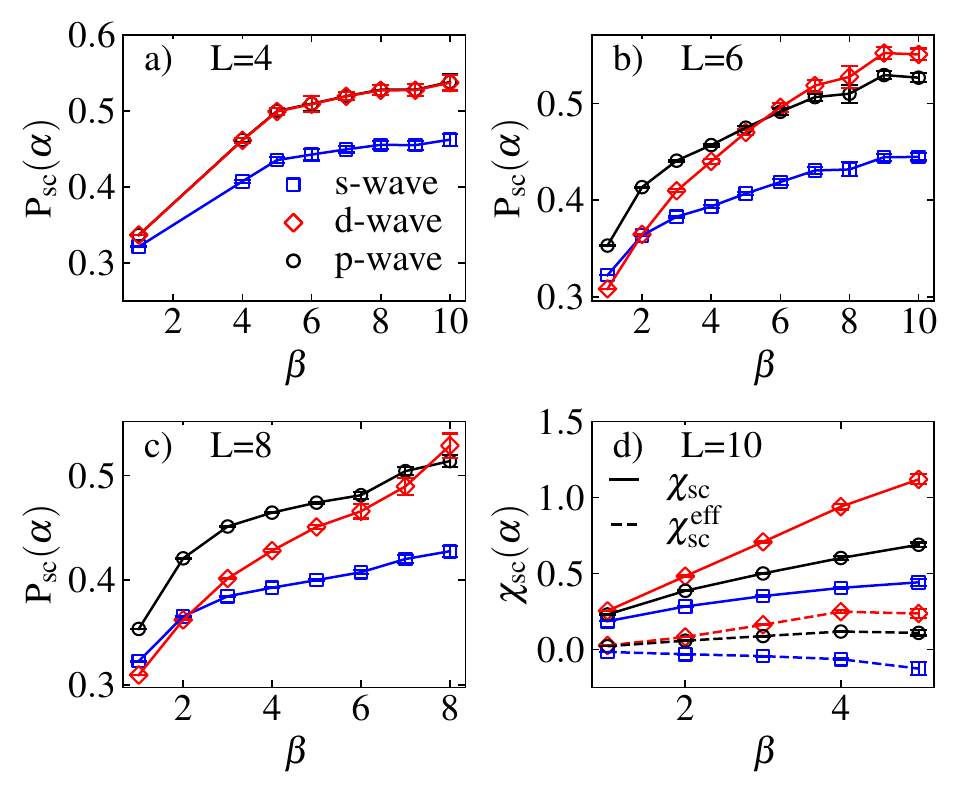}
    \caption{ (a)-(c) Pairing structure factors and (d) pairing susceptibilities as functions of the inverse temperature for different  symmetries, for the sizes shown.
    All data are for $U=0.5$, and $V=-0.5$.
    }
    \label{fig:Psc}
\end{figure}

In view of the recent suggestion that a $p$-wave SC state could be stabilised \cite{chen2022}, we examined pairing structure factors and susceptibilities. 
Typical data are shown in Fig.\,\ref{fig:Psc}: while for $L=4$, $d$-wave and $p$-wave structure factors are degenerate, and dominate over $s$-wave, for larger systems $d$-wave pairing become dominant as $\beta~\to~\infty$. 
This tendency is confirmed by the behaviour of the pairing susceptibilities (both bare and effective), for $L=10$; see Fig.\,\ref{fig:Psc}(d) As discussed in Ref.\,\onlinecite{Huang2013},  $d$-wave pairing should dominate in the $V<0$ region, due to the nesting of the Fermi surface, while the $p$-wave would be favored in the absence of such feature.
Interestingly, this is somehow observed in our results of Fig.\,\ref{fig:Psc}: the smaller the system size is, the weaker nesting effects are, leading to a spurious $p$-wave enhancement.

%%%%%%%%%%%%%%%%%%%%%%%%%%%%%%%%%%%%%%%%%%%%%%% Fig 12 %%%%%%%%%%%%%%%
%%%%%%%%%%%%%%%%%%%%%%%%%%%%%%%%%%%%%%%%%%%%%

\subsection{Ground state phase diagram}
\label{ssec:GSdiagram}

%%%%%%%%%%%%%%%%%%%%%%%%%%%%%%%%%%%%%%%%%%%%%%% Fig 10 %%%%%%%%%%%%%%%
\begin{figure}[t]
    \centering
    \includegraphics[scale=0.5]{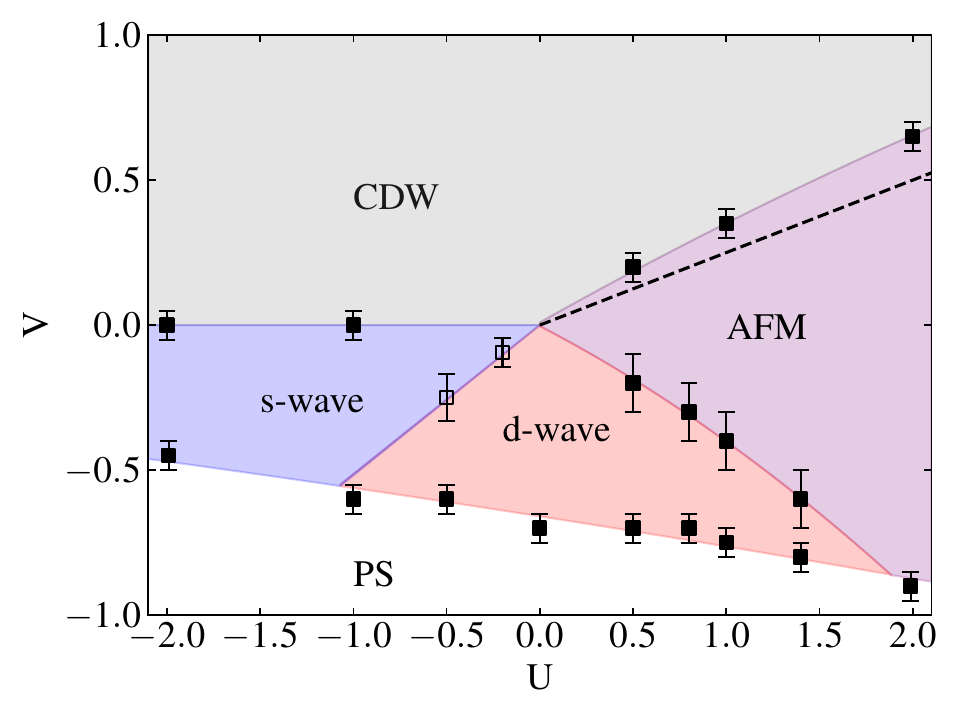}
    \caption{Ground state phase diagram for the extended Hubbard model at half filling, obtained through our DQMC simulations.
    Filled symbols are estimates for critical points obtained through the joint analyses of structure factors and $\avs$; 
    the latter played no part in determining the critical points, which are represented by empty symbols. 
    The strong-coupling AFM-CDW critical curve, $V=U/4$, is represented by a dashed line.
    Lines through data points are guides to the eye. 
    }
    \label{fig:phasediag}
\end{figure}
%%%%%%%%%%%%%%%%%%%%%%%%%%%%%%%%%%%%%%%%%%%%%%%%%%%%%%%%%%%%%%%%%%

Our findings for the ground state properties are summarized in the phase diagram of Fig.\,\ref{fig:phasediag}. 
As discussed before, the phase boundaries have been determined by examining several observables, including, in most cases, $\avs$ (calculated with real HS fields); one notable exception is the transition between $s$- and $d$-wave, which bears no signature in $\avs$. 

While previous QMC studies \cite{zhang89,Sushchyev2022,Yao22} were unable to provide a reasonably accurate critical line for the AFM-CDW transition, here we have unequivocally located the transition line above the strong coupling estimate, $V=U/4$. 
As far as the superconducting regions of the diagram are concerned, we have  set more stringent bounds for the critical points, including the transition line between $s$- and $d$-wave pairings. 
In addition, we have found no evidence of $p$-wave pairing symmetry being stabilized for any choice of parameters. 
The transition curve to the PS state has now been accurately determined over both $U<0$ and $U>0$ sectors; previous QMC estimates \cite{Yao22} were restricted to the sign-free region, $|V|\leq |U|/4$.
Still with respect to the PS boundary, it is worth stressing that an analysis for $U=2$, similar to that of Fig.\,\ref{fig:avesL6}, yields a different behavior. 
First, unlike Fig.\,\ref{fig:avesL6}(a), we have found no decrease of $S_\text{afm}$ with increasing $\beta$, characteristic of AFM being suppressed in favor of superconductivity. 
Secondly, there is only one dip in $\avs$, instead of the two dips shown in Fig.\,\ref{fig:avesL6}(c); this indicates that the SUC phase separating AFM from PS is suppressed for $U\gtrsim 2$.
Therefore, Fig.\,\ref{fig:phasediag} shows that a superconducting state in the $U>0$, $V<0$ region can only survive within a regime of intermediate couplings, namely $U\lesssim 2$ and $V\gtrsim -0.75$.

%%%%%%%%%%%%%%%%%%%%%%%% IV. Critical temperature %%%%%%%%%%%%%%%%%%%%%%%% 
\section{Critical temperatures}
\label{sec:critical_temps}

The presence of nearest-neighbor interaction breaks the degeneracy of the SUC and CDW phases for $U<0$, so that the Mermin–Wagner–Hohenberg theorem~\cite{Mermin1966,Hohenberg1967} does not apply even at half filling.
Hence, it is of interest to determine how the critical temperature for these phases changes with $V$.

%%%%%%%%%%%%%%%%%%%%%%%%%%%%%%%%%%%%%%%%%%%%%%% Fig 11 %%%%%%%%%%%%%%%
\begin{figure}[t]
    \centering    
    \includegraphics[scale=0.5]{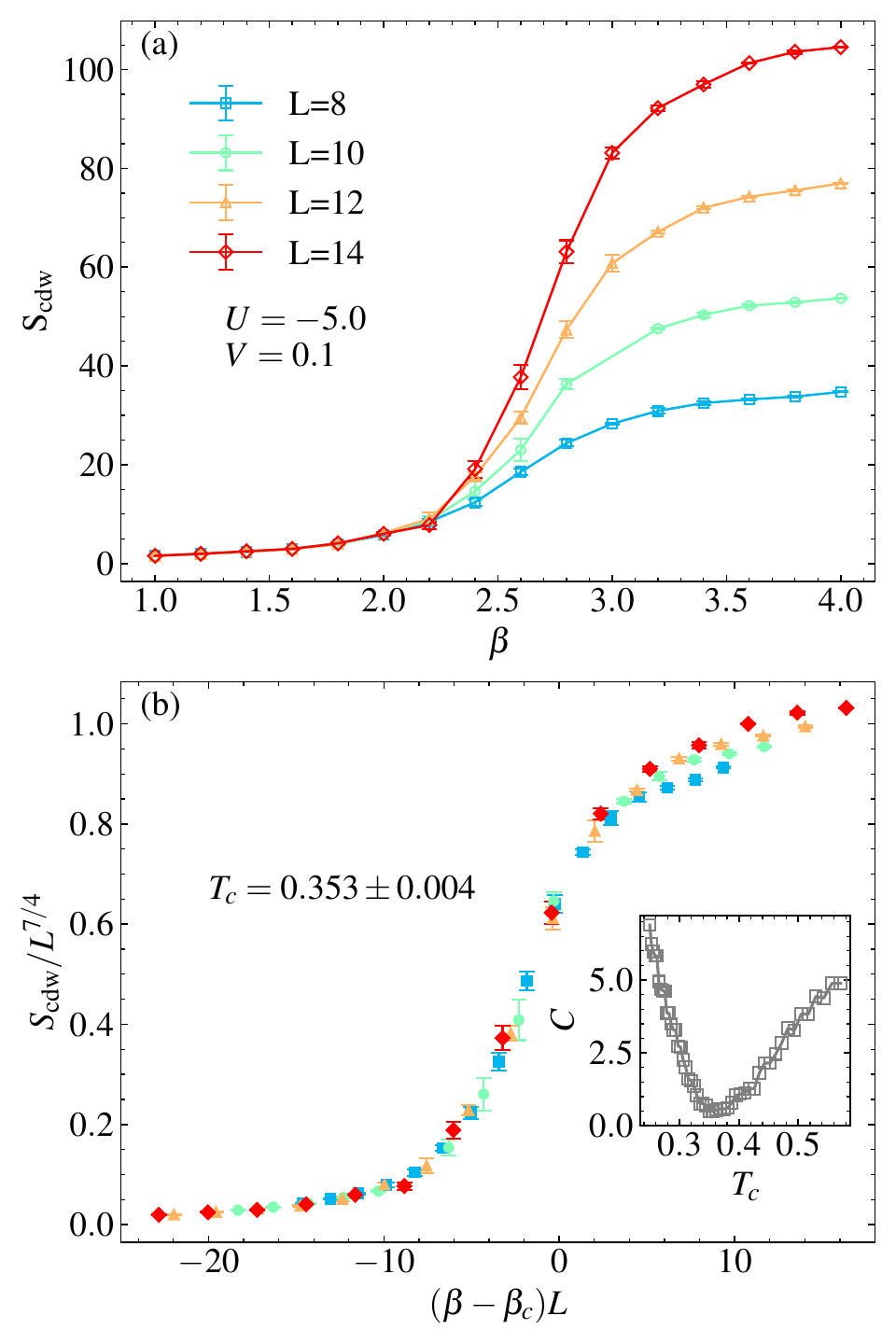}
    \caption{(a) Charge structure factor as a function of $\beta$. 
    (b) Collapse of the data in (a) to a finite-size scaling form, Eq.\,\eqref{eq:scaling_ansatz}.
    Inset: minimization of the cost function (see text) to determine $\beta_c \equiv 1/T_c$ from the data collapse.
    }
    \label{fig:tcdw}
\end{figure}
%%%%%%%%%%%%%%%%%%%%%%%%%%%%%%%%%%%%%%%%%%%%%

In order to determine the critical temperatures, we use the data for the uniform charge structure factor of Fig.\,\ref{fig:tcdw}(a) together with the finite-size scaling (FSS) ansatz \cite{Fisher71,Barber83,dosSantos81a},
\begin{equation}
    S_\text{cdw} = L^{\gamma/\nu} f[(\beta - \beta_c)L^{1/\nu}],
    \label{eq:scaling_ansatz}
\end{equation}
where, given that the temperature-driven CDW transition belongs to the two-dimensional Ising universality class, $\gamma=7/4$ and $\nu=1$; see, e.g.\, Ref.\,\cite{Stanley71}.  We may determine the critical temperature for fixed $U$ and $V$ by searching for the best fit of $S_\text{cdw}/L^{7/4}$ to Eq.\,\eqref{eq:scaling_ansatz}, namely the one minimizing the cost function $C(T_c)$, defined generically as \cite{Suntajs2020}
\begin{equation}
C(T_{c}) = \sum_{i} \frac{|f_{i+1} - f_{i}|}{\text{max}\{f_{i} \} - \text{min}\{f_{i}\}} - 1, 
\end{equation}
where $i$ runs over the set of data, and the $f_{i}$'s are the scaled structure factors, $S_\text{cdw}/L^{7/4}$, ordered according to the respective $(\beta - \beta_c)L$ values. 
When the best collapse is achieved, the distances between consecutive points are reduced, and the value of $C$ is minimized, as shown in the inset of Fig.\,\ref{fig:tcdw}(b).

\begin{figure}[t]
    \centering    
    \includegraphics[scale=0.465]{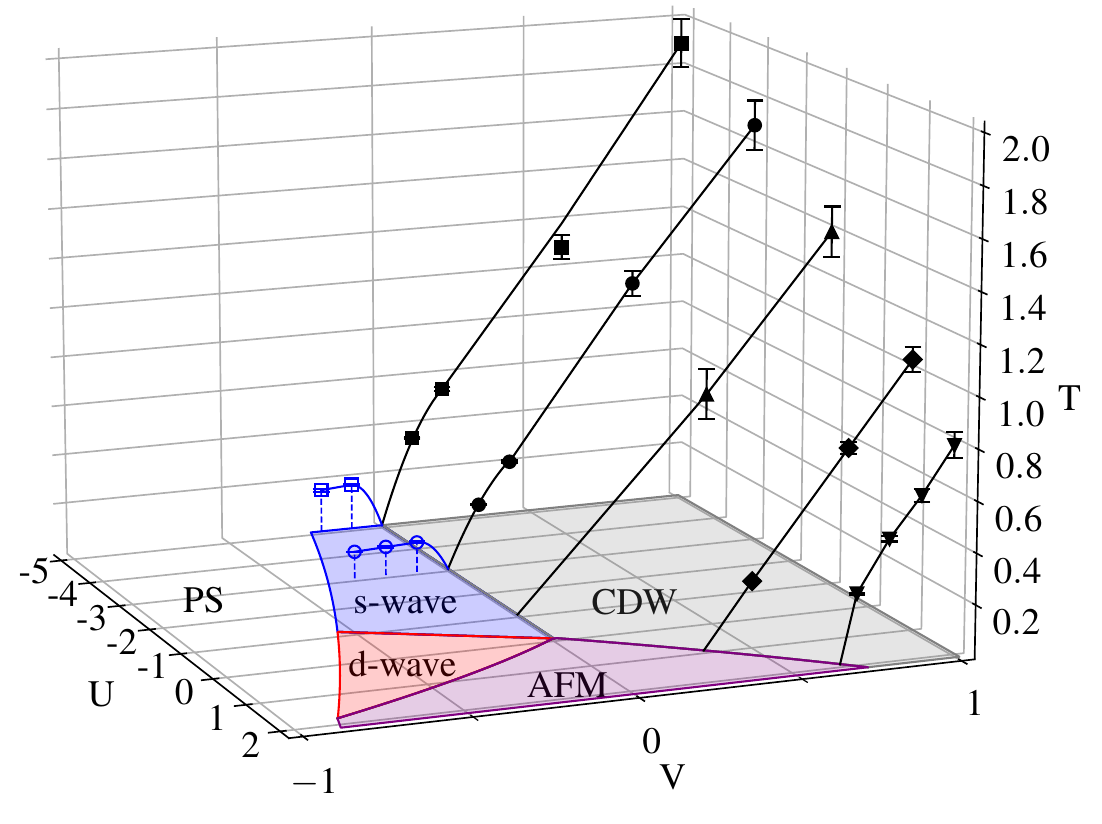}
    \caption{Critical temperatures, $T_c$, as functions of $U$ and $V$: Filled symbols correspond to $T_c$ for CDW, and empty symbols to superconducting $T_c$.
    When not shown, error bars are smaller than data points. Lines through data points are guides to the eye.
    }
    \label{fig:tcs}
\end{figure}

The behavior of $T_c^\text{CDW}$ thus obtained is shown in Fig.\,\ref{fig:tcs}.
We see that the extended interaction leads to a sharp increase in $T_c$ with increasing $V$, in the whole range of $(U,V)$ corresponding to a CDW ground state.
The quadrant $U<0,V>0$, displays $T_c$'s higher than in the first quadrant, since on-site pairs tend to be formed, and even a small $V>0$ favors double occupation of sites in one of the sublattices.
When $U>0$, on the other hand, the tendency to form an AFM state must be overcome by the nearest-neighbor repulsion.
It is also interesting to note that in the EHM one reaches critical temperatures higher than in other conventional models describing charge ordering, such as the Holstein model \cite{feng2020};
that is, the extended interaction works more efficiently, since there is no need to excite phonon modes.

%%%%%%%%%%%%%%%%%%%%%%%%%%%%%%%%%%%%%%%%%%%%%%% Fig 12 %%%%%%%%%%%%%%%
%%%%%%%%%%%%%%%%%%%%%%%%%%%%%%%%%%%%%%%%%%%%%

%%%%%%%%%%%%%%%%%%%%%%%%%%%%%%%%%%%%%%%%%%%%%%% Fig 13 %%%%%%%%%%%%%%%
\begin{figure}[t]
    \centering    
    \includegraphics[scale=0.5]{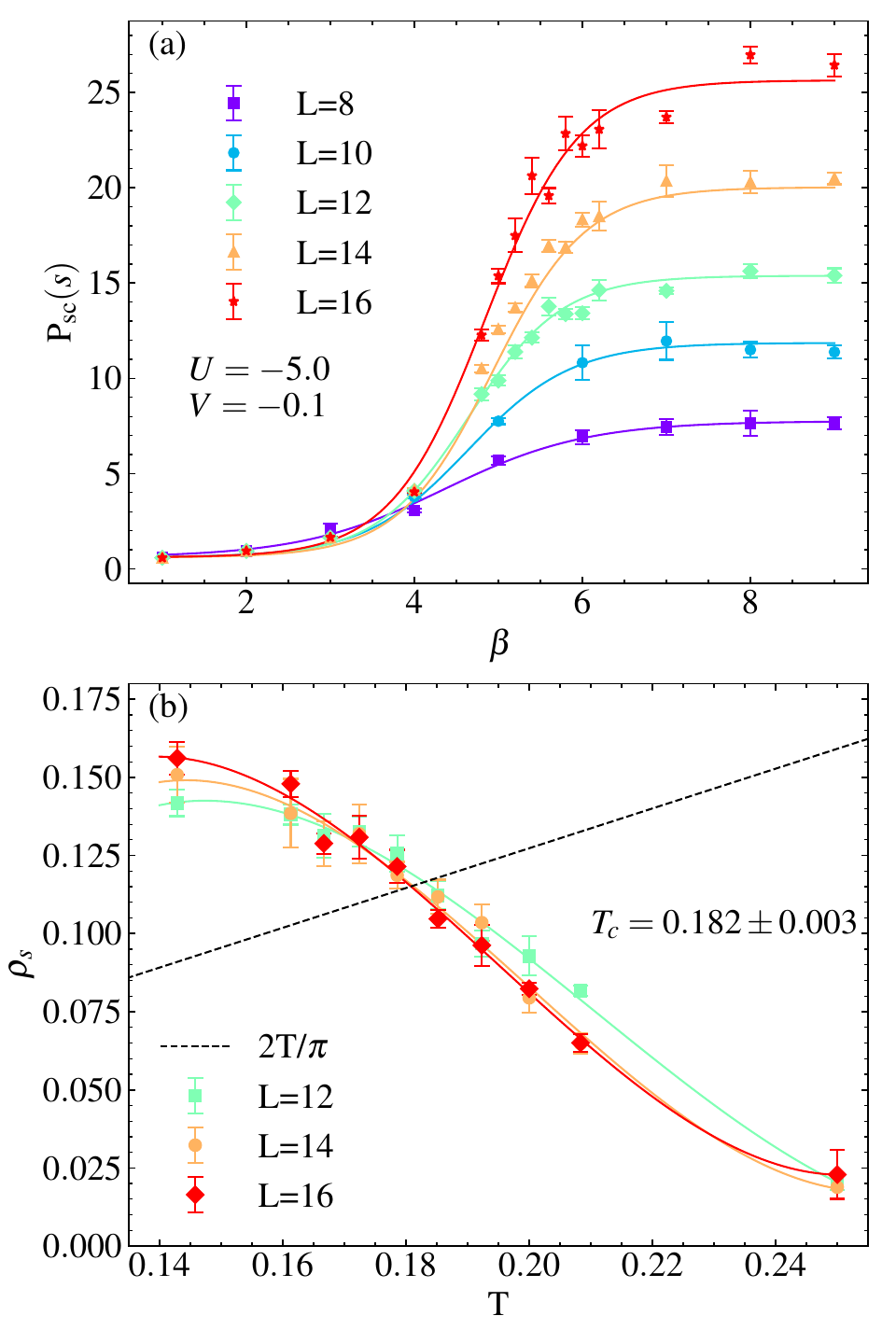}
    \caption{a) $s$-wave pairing structure factor as a function of $\beta$. b) Superfluid density as a function of temperature for different lattice sizes, the dashed line stems for the $2T/\pi$ curve.
    }
    \label{fig:superfluid}
\end{figure}
%%%%%%%%%%%%%%%%%%%%%%%%%%%%%%%%%%%%%%%%%%%%%

Let us now discuss the critical temperature for superconductivity in the third quadrant of Fig.\,\ref{fig:phasediag}.
In particular, we recall that the degraded $\avs$ (see Fig.\,\ref{fig:sign}) can be avoided if of uses CHSF in the region $|V| \leq |U|/4$.
Indeed, Fig.\,\ref{fig:superfluid}(a) shows the inverse temperature dependence of the uniform $s$-wave pairing structure factor, and, similarly to Fig.\,\ref{fig:tcdw}, the steady increase with $L$ signals a phase transition.
Estimates of $T_c$ for two-dimensional superconductivity are more efficiently determined through the superfluid density \cite{Nelson77,Scalapino92,Scalapino93},
\begin{equation}
    \rho_s = \frac{1}{4} \left[-K_x -\Lambda_{xx}(q_x=0,q_y\to 0,\omega=0) \right],
\end{equation}
where $K_x$ is the kinetic energy for motion along the $x$ direction, and $\Lambda_{xx}$ is the current correlator \cite{Scalapino92,Scalapino93}.
The superconducting critical temperature $T_c$ is calculated with the aid of the jump discontinuity \cite{Nelson77},
\begin{equation}
    T_c = \frac{\pi}{2} \rho_s^-, 
\end{equation}
where $\rho_s^-$ is the value of the superfluid density just below the critical temperature. 

In Fig.\,\ref{fig:superfluid}(b) we plot $\rho_s$ as a function of $T$, for fixed $U$, $V$, and $L$, and the intercept with the straight line $2T/\pi$ provides an estimate for $T_c$. 
We see that the intercepts are very weakly dependent on $L$, so that finite-size effects are mitigated \cite{Paiva04,fontenele2022}.
Figure \ref{fig:tcs} shows the trends of superconducting $T_c$ for other values of $U$ and $V$.
As $V$ decreases, $T_c$ initially grows, but tends to saturate, and we recall that further decrease in $V$ drives the system to a PS regime, hence the interrupted curves.
Overall, for fixed $V$, $T_c$ tends to increase with increasing $|U|$, but the superconducting region is eventually suppressed in favor of PS. 
It is also noteworthy that these $V$-enhanced critical temperatures are even higher than the maximum value obtained for $V=0$ \cite{fontenele2022}.  

We have also analyzed the $d$-wave region, starting with an FSS ansatz for the pairing structure factor \cite{dosSantos93}, according to which a Kosterlitz-Thouless transition is signaled by a crossing of the curves for different sizes.   
In contrast with the third quadrant, no crossing was found for $\beta < 10$, with $V=-0.5$ and for two values of $U$, namely $U=0$ and $U=0.5$. 
Unfortunately, the minus sign problem is much more severe in this region, which precludes any analysis for $\beta>10$; note that this region lies outside the sign-free region through CHSF.
Nonetheless, one may conclude that $T_c$ is much lower than those found in the third quadrant for $s$-wave pairing.
This, in turn, allows one to conjecture an exponential behavior with $V<0$ for fixed $U\geq0$, i.e.\ $T_c\sim \exp(-C/|V|)$, where $C$ is independent of $V$; an alternative linear dependence with $|V|$ emerged from a recent mean-field approach \cite{Sun24}.
It is interesting to note that if one takes order of magnitude data from the cuprates, such as a bandwidth $W\sim 10$ eV \cite{Mattheiss87} and $T_c\sim 10^2$K, we get (in the dimensioless units used here) $\beta_c \sim 10^2$, a range beyond our limitations imposed by the sign problem.

%%%%%%%%%%%% V. Conclusions %%%%%%%%%%%%%%%
\section{Conclusions}
\label{sec:conclusions}

In summary, we have resolved long standing issues relative to the ground state phase diagram of the half-filled extended Hubbard model on a square lattice, such as the pairing symmetries of the superconducting phase, and locations of phase boundaries in all four quadrants of the $U-V$ plane. 
Indeed, through our DQMC simulations, accurate boundaries involving antiferromagnetic, charge-density wave, $s$-wave and $d$-wave superconducting, and phase-separated phases were determined.
It is worth emphasizing that the `minus-sign' problem of QMC simulations was overcome by performing extensive simulations, and by using the recently proposed connection between critical points and strong dips in $\avs$.
Indeed, the multitude of phases in the diagram allowed us to verify that dips in $\avs$ only occur at transitions involving different universality classes, such as AFM-CDW, AFM-SC, SC-PS, and CDW-SC, but not between $s$- and $d$-waves.

We have also determined the critical temperatures for the CDW and $s$-wave superconducting phases. For the CDW phase, we found that for fixed $U$ the critical temperature increases sharply with $V$, reaching higher $T_c$'s  than in other electronic models for CDW's such as the Holstein model. 
Hopefully these findings will stimulate experiments with ultracold atoms interacting beyond on-site couplings: phase transitions at temperatures within a feasible range could be probed with a quantum gas microscope.

The presence of a $d$-wave superconducting ground state over a reasonably wide region of the parameter space has bearings on the high-$T_c$ cuprates, particularly in the $U>0$, $V<0$ region.  
Our analyses of the finite temperature data for the $d$-wave structure factor suggests $\beta_c$ somewhat larger than 10 (in dimensionless units), which is consistent  with actual data for the cuprates; note that the $T_c$'s predicted here for the $s$-wave pairing in the $U<0$, $V<0$ region are much higher than those for $d$-wave.
These results add credence to the use of the EHM as a minimal single-band model (with $U>0$ and $V<0$) to describe the high-$T_c$ cuprates\,\cite{Scalapino12}. 
Nonetheless, a more stringent test would be to investigate the properties of the current in the doped regime.

%%%%%%%%%%%%%%%%%%%%%%%%%%%%%%%%%%%%%%%%%%%%%%%%%%%%%%%%%%%%%%%%%%
%%%%%%%%%%%%%%%%%%%%%%%%  ACKNOWLEDGMENTS  %%%%%%%%%%%%%%%%%%%%%%%
%%%%%%%%%%%%%%%%%%%%%%%%%%%%%%%%%%%%%%%%%%%%%%%%%%%%%%%%%%%%%%%%%%
\section*{ACKNOWLEDGMENTS}
The authors are grateful to the Brazilian Agencies Conselho Nacional de Desenvolvimento Cient\'\i fico e Tecnol\'ogico (CNPq), Coordena\c c\~ao de Aperfei\c coamento de Pessoal de Ensino Superior (CAPES),  and Instituto Nacional de Ci\^encia e Tecnologia de Informa\c c\~ao Qu\^antica (INCT-IQ) for funding this project.
N.C.C.~acknowledges support from FAPERJ Grant No.~E-26/200.258/2023 - SEI-260003/000623/2023, and CNPq Grant No.~313065/2021-7.

%%%%%%%%%%%%%%%%%%%%%%%%%%%%%%%%%%%%%%%%%%%%%%%%%%%
%%%%%%%%%%%%%%%%%%%  Appendices  %%%%%%%%%%%%%%%%%%
%%%%%%%%%%%%%%%%%%%%%%%%%%%%%%%%%%%%%%%%%%%%%%%%%%%
\appendix

\section{Real Hubbard-Stratonovich fields (RHSF)}
\label{App:RHSF}
In preparing for the simulations, the interaction terms are separated through a Suzuki-Trotter decomposition, 
\begin{equation}
    e^{-\beta\mathcal{H}} \approx \left( e^{-\Delta\tau\mathcal{H}_k}e^{-\Delta\tau\mathcal{H}_U}e^{-\Delta\tau\mathcal{H}_V} \right)^{L_\tau} +\mathcal{O}(\Delta\tau^2),
\end{equation}
so that the 2D problem is mapped onto a 3D system, with the extra dimension being the imaginary time axis, and the inverse of the temperature $\beta$ is cut into $L_\tau$ discrete intervals with length $\Delta\tau = \beta/L_{\tau}$. 
The quartic terms in $\mathcal{H}_U$ and $\mathcal{H}_V$ are expressed in a quadratic form using the Hubbard-Stratonovich transformation \cite{Hirsch83,zhang89},
\begin{equation}
    e^{-\Delta\tau W n_{\iv\s}n_{\jv\s^\prime } }= \frac{1}{2} \sum_{x_\nu=\pm 1} e^{ \left[ \alpha x_\nu (n_{\iv\s} - n_{\jv\s^\prime }) -\frac{\Delta\tau W}{2}  (n_{\iv\s} + n_{\jv\s^\prime }) \right]}
    \label{hst}
\end{equation}
with $\cosh{\alpha } = \exp{\Delta\tau W / 2}$, and $W$ is either $U$ or $V$, respectively if $\jv=\iv$ or if $\jv$ is a first neighbor of $\iv$; $\s$ and $\s'$ stand for the original fermionic spin variable, $\uparrow$ or $\downarrow$.
The second term in the argument of the exponential in Eq.\,\eqref{hst} vanishes at half filling, that is, for $\mu = U/2 + 4V$. 
When dealing with the on-site coupling, $\mathcal{H}_U$, we define one \emph{species} of auxiliary field, $x_0(\iv, l)$, at each lattice site $\iv$ and time slice $l$. 
For the nearest-neighbor coupling, $\mathcal{H}_V$, we write
\begin{equation}
    V n_{\iv}n_{\jv} = V(  n_{\iv\uparrow}n_{\jv\uparrow }+n_{\iv\uparrow}n_{\jv\downarrow }+n_{\iv\downarrow}n_{\jv\uparrow }+n_{\iv\downarrow}n_{\jv\downarrow } ),
\end{equation}
and define four additional species of auxiliary fields for each bond between sites $\iv$ and $\jv$, namely $x_1(\iv,\jv, l)$, $x_2(\iv,\jv, l)$, $x_3(\iv,\jv, l)$, and $x_4(\iv,\jv, l)$.
For an $N$-site square lattice with periodic boundary conditions, there are $2 \times N$ bonds, hence $8 \times N \times L_\tau$ auxiliary fields, in addition to the $N \times L_\tau$ auxiliary fields for the on-site term. 
The partition function then becomes,
\begin{equation}
    \mathcal{Z} = \dfrac{1}{2^{9NL_\tau}} \sum_{x_\nu=\pm 1} \prod_\sigma\det{A^\sigma(x_0,x_1,x_2,x_3,x_4) },
\end{equation}
with,
\begin{equation}
    A^\sigma(x_0,x_1,x_2,x_3,x_4) = \mathbb{1} + B^\sigma (L_\tau)B^\sigma (L_\tau-1) \cdots B^\sigma (1) 
\end{equation}
and
\begin{equation}
    B^\sigma (l) =e^{-\Delta\tau\hat{H}_k}e^{-\Delta\tau\hat{H}_U}e^{-\Delta\tau\hat{H}_{V}}. 
\end{equation}
The weight $P(x_\nu)$ of each configuration of auxiliary fields is therefore given by
\begin{equation}
    P(x_\nu) = \left|\det \left[ A^\uparrow A^\downarrow \right] \right|.
\end{equation}
One can also compute Green's functions \cite{Rademaker2013},
\begin{equation}
    G_{\iv,\jv}^\s = \dfrac{1}{2^{9NL_\tau}} \sum_{x_\nu=\pm 1} \left[ A^\s\right]^{-1}_{\iv,\jv} \prod_\sigma\det{A^\sigma(x_0,x_1,\cdots,x_4) },
    \label{grnf}
\end{equation}
and calculate the relevant physical quantities. 
The simulations are then carried out by importance-sampling the $2^{9NL_\tau}$ Hubbard-Stratonovich fields (HSF), $\{x_\nu \}$, taken as  Ising variables, $\pm1$.  

\section{Complex Hubbard-Stratonovich fields (CHSF)}
\label{App:CHSF}

The use of complex Hubbard-Stratonovich fields (CHSF) emerges as an alternative to mitigate the minus sign problem \cite{golor2015,Yao22}.
We start by defining an interaction term as
\begin{equation}
\mathcal{H}_I=\frac{g}{2}\sum_{\ave{\iv,\jv}} \Lambda_{\iv\jv} ^2,
\end{equation}
where for each bond between sites $\iv$ and $\jv$ we define
$\Lambda_{\iv\jv} \equiv n_\iv-1 +a(n_\jv-1)$,
with $a=V/g$, and
$g=U/(2+2a^2)$.
In order to recover the original interaction terms in Eq.\,\eqref{eq:EHM}, $a$ must be a solution of the quadratic equation,
$2Va^2 -Ua+2V =0$, which only admits a real solution if $|V|\leq |U|/4$.
Thus the introduction of CHSF is restricted to this range.

One may then write
\begin{equation}
    e^{-\Delta\tau g\Lambda_{\iv\jv}^2/2} \approx \sum_{\nu=\pm1,\pm2} \eta(\nu)\ e^{x_{\nu} \sqrt{-g\Delta\tau} \Lambda_{\iv\jv}},
\end{equation}
with $\eta(\pm 1) \equiv (1+\sqrt{6}/3)/4$, $\eta(\pm 2) \equiv (1-\sqrt{6}/3)/4$,
$x_{\pm 1} \equiv\pm \sqrt{3-\sqrt{6}}$, and $x_{\pm 2} \equiv \pm\sqrt{3+\sqrt{6}}$.
One should keep in mind that when dealing with RHSF, each species can be in two (Ising) states, while with CHSF one has only one species, which can be in four states, $\nu=\pm1,\pm2$. 
It is worth mentioning that this transformation is not exact, since it introduces an error of $\mathcal{O}(\Delta\tau^4)$; nonetheless, this is negligible in comparison with the Trotter error, which is $\mathcal{O}(\Delta\tau^2)$. 
We also note that when $g<0$, the  argument of the exponential on the RHS is real, so that the up- and down-determinants are equal. 
Further, if $g>0$ the exponential yields a complex number, but if   particle-hole symmetry is satisfied, the determinants are complex conjugate of each other, so that their product is positive definite \cite{golor2015,Yao22}, and the simulation is sign-free.

In the formulation with CHSF, Eq.\,\eqref{grnf} is replaced by 
\begin{equation}
    G_{\iv,\jv}^\s = \dfrac{1}{2^{2NL_\tau}} \sum_{\nu=\pm 1,\pm 2} \left[ \Lambda^\s\right]^{-1}_{\iv,\jv} \prod_\sigma\det{A^\sigma(x_{\nu}) },
    \label{grnf2}
\end{equation}
Now, the number of Hubbard-Stratonovich fields grows as $2^{2NL_\tau}$, which renders the code even faster in comparison with the (RHSF) case within the sign-free region $|V|\leq |U|/4$.

\section{Sampling the Hubbard-Stratonovich fields:}

The Ising sampling of the HSF consists of sweeping over the sites and bonds in each imaginary time slice $l$, and attempting to flip $x_\nu \rightarrow -x_\nu $. 
Say a flip of the $x_0$ field at a single site on a time slice is proposed: if the change is accepted, the new Green's function is computed through $O(N^2)$ operations using the Sherman-Morrison update \cite{gubernatis2016}, instead of computing it from scratch using Eq.\,\eqref{grnf}, which requires $O(N^3)$ operations. 
The same approach can be used to update the Green's functions by flipping the $x_2$ and $x_3$ fields, which couples the terms with $\s \neq \s^\prime $ in Eq.\,\eqref{hst}, and consequently allows us to use Sherman-Morrison update for $G^\uparrow$ and $G^{\downarrow}$ separately. 
The challenge is to update the Green's functions by flipping the $x_1$ and $x_4$ fields since they change the Green's functions related to different sites for $\s = \s^\prime$. Following the procedure described in \cite{Rademaker2013} we use the Woodbury matrix identity, which is a generalized Sherman-Morrison update. The steps to implement this update are detailed below.
\begin{figure*}[t]
    \centering
    \includegraphics[scale=0.5]{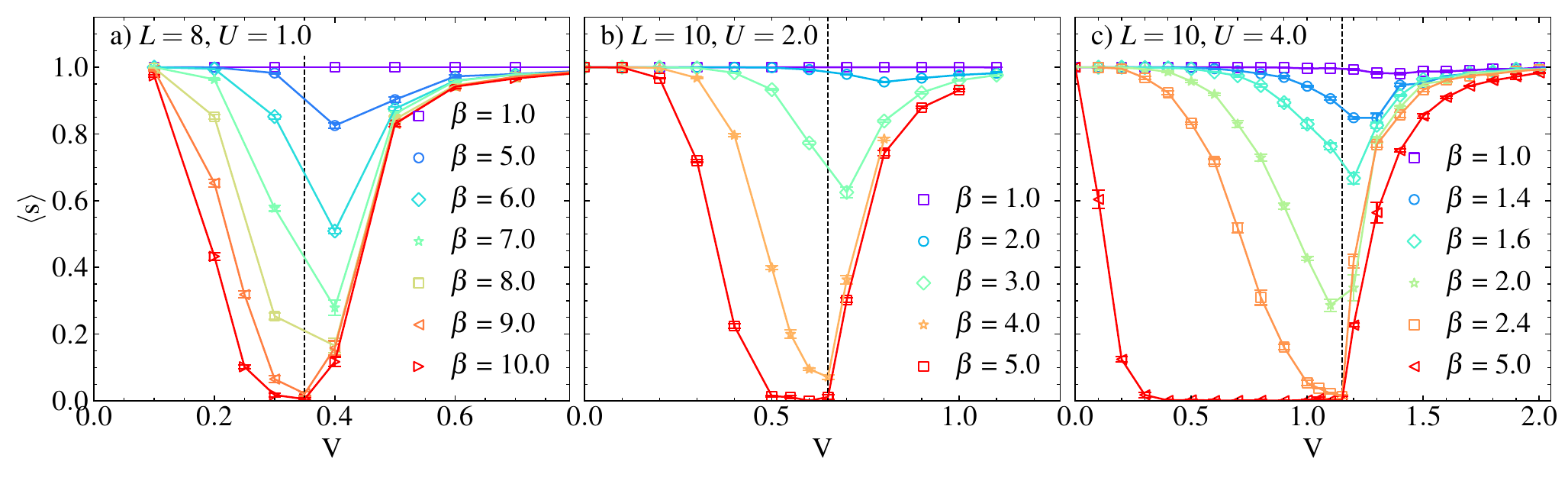}
    \caption{Average fermion sign as a function of $V$ for (a) $L=8$ and $U=1.0$, (b)$L=10$ and $U=2.0$ and (c)$L=10$ and $U=4.0$. The dashed lines indicates the estimated critical values of $V$.}
    \label{fig:signL10}
\end{figure*}
Starting with a known Green's function at a time slice $l$, and given the fields $x_{1(4)}$, one proposes a flip $x_{1(4)} \rightarrow x_{1(4)}^\prime = -x_{1(4)} $ in a bond between sites $\iv$ and $\jv$ for a time slice $l$ individually. Under this change the $A^\s$ matrix becomes,
\begin{equation}
    A^\s  \rightarrow \mathbb{1} - B^\sigma (L_\tau)\cdots \left( \mathbb{1}  +\Delta^\s \right) B^\s(l) \cdots B^\sigma (1) = A^{\prime \s}.   
\end{equation}
The label $\s$ corresponds to spin up or down for $x_1$ and $x_4$ respectively. The matrix $\Delta^\s$ is diagonal and has only two nonzero elements, namely
\begin{equation}
\begin{split}
    \Delta_{i,i}^\s & = \exp\left[-2 x_\nu(\iv,\jv,l) \right] -1\\
    \Delta_{j,j}^\s &= \exp\left[ 2 x_\nu(\iv,\jv,l) \right] -1.
\end{split}
\end{equation}
One can decide whether or not to accept a change using the ratio of determinants \cite{gubernatis2016},
\begin{equation}
    R^\s = \dfrac{\det \left[ A^{\prime\s} \right] }{\det \left[ A^\s \right] } = \det \left[ \mathbb{1} - \Delta^\s( \mathbb{1} - G^\s ) \right].
    \label{eq:detrattot}
\end{equation}
If the flipping is accepted, by using matrix identities it is possible to write a simple expression for the updated $G^{\s\prime}$ after the flip \cite{gubernatis2016},
\begin{equation}
    G^{\prime\s} \rightarrow G^{\s} - G^{\s}\dfrac{1}{ \mathcal{R} } \Delta^\s (\mathbb{1} - G^\s).
    \label{eq:woodburyfu}
\end{equation}
Due to the sparseness of the matrices $\mathcal{R}$ and $\Delta^\s$, Eq.\,\eqref{eq:woodburyfu} can be written in terms of matrix elements \cite{Rademaker2013},
\begin{equation}
    G^{\prime\s}_{r,s} \rightarrow G^{\s}_{r,s} - \sum_{i,j} G^{\s}_{r,i}  \mathcal{D}_{i,j} (\delta_{j,s} - G^\s_{j,s}),
\end{equation}
where the matrix $\mathcal{D}$ may be cast in a $2\times 2$ form,
\begin{equation}
    \mathcal{D} = \dfrac{\Delta^\s_{i,i} \Delta^\s_{j,j}}{R^\s} \begin{pmatrix}
         1- G_{j,j}^\s +\dfrac{1}{\Delta^\s_{j,j}} &  G^{\s}_{i,j} \\
        & \\
     G^{\s}_{j,i} & 1-G_{i,i}^\s + \dfrac{1}{\Delta^\s_{i,i}}
    \end{pmatrix}
\end{equation}
with $i$ and $j$ being the indices of the sites related with the flip of $x_\nu(\iv,\jv,l)$, and, 
\begin{equation}
\begin{split}
     R^\s= \det \mathcal{R}  = & [\Delta^\s_{i,i} (1-G_{i,i}^\s) + 1] [\Delta^\s_{j,j}(1- G_{j,j}^\s) +1]\\
    & - \Delta^\s_{i,i} G_{i,j}^\s  \Delta^\s_{j,j} G_{j,i}^\s.
\end{split}
\end{equation}
In order to decide whether a proposed change is accepted we use a combination of the Metropolis and the heat-bath algorithm, similar to the one used in Ref.\,\cite{Rademaker2013}, 
\begin{equation}
  P_T=  \left\{
    \begin{array}{ll}
      \dfrac{R^\s}{1+\gamma R^\s} 
      & \text{if } R^\s \leq 1\\
      & \\
      \dfrac{R^\s}{\gamma + R^\s} 
      & \text{if } R^\s > 1
    \end{array}
  \right.  
\end{equation}
The parameter $\gamma$ is tuned self-consistently to achieve an acceptance ratio of approximated $50\%$.

The order of flipping attempts is also a crucial detail in the DQMC method when more than one auxiliary field per site/bond. 
We have tried four different strategies. 
As it turned out, the most efficient is the following. 
A sweep attempting to flip just $x_0(\iv,l)$ is carried out over all sites and all time slices, $(\iv,l)$. 
This is followed by a sweep over all \emph{bonds} and time slices, trying to flip just $x_1(\iv,\jv,l)$, then followed by another sweep over the whole space-time lattice attempting to flip just $x_2(\iv,\jv,l)$, and so on and so forth for $x_3(\iv,\jv,l)$ and for $x_4(\iv,\jv,l)$ to complete one Monte Carlo step.
Any attempt of grouping more than one flip at any $x_i$, despite speeding up the sweep, leads to much more noisy averages than with the above mentioned strategy.

As a final remark, we note that when dealing with CHSF, one is restricted to Metropolis algorithm, since the heat bath algorithm leads to a very low acceptance ratio.

\section{Asymptotic behavior of the fermion sign for the AFM-CDW transition.}
\label{App:sign}

Figure \ref{fig:avesB4} in the main text shows that the fermion sign exhibits a dip near the critical value of $V$ at a fixed temperature. 
Here we extend the analysis a bit further, by discussing the behavior of $\avs$ as $\beta\to\infty$.  
Figure \ref{fig:signL10} shows data for $\avs$ for three different values of $U$, near their respective AFM-CDW critical points. 
Starting with Fig.\,\ref{fig:signL10}(a), we see that as $V$ decreases from the CDW phase, the dip in $\avs$ deepens considerably as $\beta$ increases, such that $\avs\to0$ at $V\approx 0.35$, as indicated by the dotted vertical line; this would be even more apparent in a log-linear scale. 
Therefore, estimates obtained from $\avs$ at higher temperatures (Sec.\,\ref{ssec:AFM-CDW}) are not strongly dependent on $\beta$.
Indeed, Fig.\,\ref{fig:signL10}(b) complements Fig.\,\ref{fig:avesB4}, in the sense that now $L$ is fixed, and $\beta$ is increased: $\avs\to0$ at $V\approx0.65$, again within the error bars quoted in Sec.\,\ref{ssec:AFM-CDW}.
Data for $U=4$ displayed in Fig.\,\ref{fig:signL10}(c) provides yet another consistency check that $\avs\to0$ signals a critical point.

%%%%%%%%%%%%%%%%%%%%%%%%%%%%%%%%%%%%%%%%%%%%%
%%%%%%%%%%%%    bibliography   %%%%%%%%%%%%%%
%%%%%%%%%%%%%%%%%%%%%%%%%%%%%%%%%%%%%%%%%%%%%
\bibliography{ref}

%apsrev4-2.bst 2019-01-14 (MD) hand-edited version of apsrev4-1.bst
%Control: key (0)
%Control: author (72) initials jnrlst
%Control: editor formatted (1) identically to author
%Control: production of article title (-1) disabled
%Control: page (0) single
%Control: year (1) truncated
%Control: production of eprint (0) enabled
\begin{thebibliography}{56}%
\makeatletter
\providecommand \@ifxundefined [1]{%
 \@ifx{#1\undefined}
}%
\providecommand \@ifnum [1]{%
 \ifnum #1\expandafter \@firstoftwo
 \else \expandafter \@secondoftwo
 \fi
}%
\providecommand \@ifx [1]{%
 \ifx #1\expandafter \@firstoftwo
 \else \expandafter \@secondoftwo
 \fi
}%
\providecommand \natexlab [1]{#1}%
\providecommand \enquote  [1]{``#1''}%
\providecommand \bibnamefont  [1]{#1}%
\providecommand \bibfnamefont [1]{#1}%
\providecommand \citenamefont [1]{#1}%
\providecommand \href@noop [0]{\@secondoftwo}%
\providecommand \href [0]{\begingroup \@sanitize@url \@href}%
\providecommand \@href[1]{\@@startlink{#1}\@@href}%
\providecommand \@@href[1]{\endgroup#1\@@endlink}%
\providecommand \@sanitize@url [0]{\catcode `\\12\catcode `\$12\catcode `\&12\catcode `\#12\catcode `\^12\catcode `\_12\catcode `\%12\relax}%
\providecommand \@@startlink[1]{}%
\providecommand \@@endlink[0]{}%
\providecommand \url  [0]{\begingroup\@sanitize@url \@url }%
\providecommand \@url [1]{\endgroup\@href {#1}{\urlprefix }}%
\providecommand \urlprefix  [0]{URL }%
\providecommand \Eprint [0]{\href }%
\providecommand \doibase [0]{https://doi.org/}%
\providecommand \selectlanguage [0]{\@gobble}%
\providecommand \bibinfo  [0]{\@secondoftwo}%
\providecommand \bibfield  [0]{\@secondoftwo}%
\providecommand \translation [1]{[#1]}%
\providecommand \BibitemOpen [0]{}%
\providecommand \bibitemStop [0]{}%
\providecommand \bibitemNoStop [0]{.\EOS\space}%
\providecommand \EOS [0]{\spacefactor3000\relax}%
\providecommand \BibitemShut  [1]{\csname bibitem#1\endcsname}%
\let\auto@bib@innerbib\@empty
%</preamble>
\bibitem [{\citenamefont {Bednorz}\ and\ \citenamefont {M{\"u}ller}(1986)}]{Bednorz1986}%
  \BibitemOpen
  \bibfield  {author} {\bibinfo {author} {\bibfnamefont {J.~G.}\ \bibnamefont {Bednorz}}\ and\ \bibinfo {author} {\bibfnamefont {K.~A.}\ \bibnamefont {M{\"u}ller}},\ }\href {https://doi.org/10.1007/BF01303701} {\bibfield  {journal} {\bibinfo  {journal} {Zeitschrift f{\"u}r Physik B Condensed Matter}\ }\textbf {\bibinfo {volume} {64}},\ \bibinfo {pages} {189} (\bibinfo {year} {1986})}\BibitemShut {NoStop}%
\bibitem [{\citenamefont {Varma}(2020)}]{Varma2020}%
  \BibitemOpen
  \bibfield  {author} {\bibinfo {author} {\bibfnamefont {C.~M.}\ \bibnamefont {Varma}},\ }\href {https://doi.org/10.1103/RevModPhys.92.031001} {\bibfield  {journal} {\bibinfo  {journal} {Rev. Mod. Phys.}\ }\textbf {\bibinfo {volume} {92}},\ \bibinfo {pages} {031001} (\bibinfo {year} {2020})}\BibitemShut {NoStop}%
\bibitem [{\citenamefont {Zhou}\ \emph {et~al.}(2021)\citenamefont {Zhou}, \citenamefont {Lee}, \citenamefont {Imada}, \citenamefont {Trivedi}, \citenamefont {Phillips}, \citenamefont {Kee}, \citenamefont {T{\"o}rm{\"a}},\ and\ \citenamefont {Eremets}}]{Zhou2021}%
  \BibitemOpen
  \bibfield  {author} {\bibinfo {author} {\bibfnamefont {X.}~\bibnamefont {Zhou}}, \bibinfo {author} {\bibfnamefont {W.-S.}\ \bibnamefont {Lee}}, \bibinfo {author} {\bibfnamefont {M.}~\bibnamefont {Imada}}, \bibinfo {author} {\bibfnamefont {N.}~\bibnamefont {Trivedi}}, \bibinfo {author} {\bibfnamefont {P.}~\bibnamefont {Phillips}}, \bibinfo {author} {\bibfnamefont {H.-Y.}\ \bibnamefont {Kee}}, \bibinfo {author} {\bibfnamefont {P.}~\bibnamefont {T{\"o}rm{\"a}}},\ and\ \bibinfo {author} {\bibfnamefont {M.}~\bibnamefont {Eremets}},\ }\href {https://doi.org/10.1038/s42254-021-00324-3} {\bibfield  {journal} {\bibinfo  {journal} {Nature Reviews Physics}\ }\textbf {\bibinfo {volume} {3}},\ \bibinfo {pages} {462} (\bibinfo {year} {2021})}\BibitemShut {NoStop}%
\bibitem [{\citenamefont {Anderson}(1987)}]{Anderson87}%
  \BibitemOpen
  \bibfield  {author} {\bibinfo {author} {\bibfnamefont {P.~W.}\ \bibnamefont {Anderson}},\ }\href {https://doi.org/10.1126/science.235.4793.1196} {\bibfield  {journal} {\bibinfo  {journal} {Science}\ }\textbf {\bibinfo {volume} {235}},\ \bibinfo {pages} {1196} (\bibinfo {year} {1987})}\BibitemShut {NoStop}%
\bibitem [{\citenamefont {Emery}(1987)}]{Emery87}%
  \BibitemOpen
  \bibfield  {author} {\bibinfo {author} {\bibfnamefont {V.~J.}\ \bibnamefont {Emery}},\ }\href {https://doi.org/10.1103/PhysRevLett.58.2794} {\bibfield  {journal} {\bibinfo  {journal} {Phys. Rev. Lett.}\ }\textbf {\bibinfo {volume} {58}},\ \bibinfo {pages} {2794} (\bibinfo {year} {1987})}\BibitemShut {NoStop}%
\bibitem [{\citenamefont {Scalapino}(2012)}]{Scalapino12}%
  \BibitemOpen
  \bibfield  {author} {\bibinfo {author} {\bibfnamefont {D.~J.}\ \bibnamefont {Scalapino}},\ }\href {https://doi.org/10.1103/RevModPhys.84.1383} {\bibfield  {journal} {\bibinfo  {journal} {Rev. Mod. Phys.}\ }\textbf {\bibinfo {volume} {84}},\ \bibinfo {pages} {1383} (\bibinfo {year} {2012})}\BibitemShut {NoStop}%
\bibitem [{\citenamefont {Lin}\ \emph {et~al.}(1995)\citenamefont {Lin}, \citenamefont {Gagliano}, \citenamefont {Campbell}, \citenamefont {Fradkin},\ and\ \citenamefont {Gubernatis}}]{Lin1995}%
  \BibitemOpen
  \bibfield  {author} {\bibinfo {author} {\bibfnamefont {H.~Q.}\ \bibnamefont {Lin}}, \bibinfo {author} {\bibfnamefont {E.~R.}\ \bibnamefont {Gagliano}}, \bibinfo {author} {\bibfnamefont {D.~K.}\ \bibnamefont {Campbell}}, \bibinfo {author} {\bibfnamefont {E.~H.}\ \bibnamefont {Fradkin}},\ and\ \bibinfo {author} {\bibfnamefont {J.~E.}\ \bibnamefont {Gubernatis}},\ }\bibinfo {title} {The phase diagram of the one-dimensional extended {H}ubbard model},\ in\ \href {https://doi.org/10.1007/978-1-4899-1042-4_35} {\emph {\bibinfo {booktitle} {The Hubbard Model: Its Physics and Mathematical Physics}}},\ \bibinfo {editor} {edited by\ \bibinfo {editor} {\bibfnamefont {D.}~\bibnamefont {Baeriswyl}}, \bibinfo {editor} {\bibfnamefont {D.~K.}\ \bibnamefont {Campbell}}, \bibinfo {editor} {\bibfnamefont {J.~M.~P.}\ \bibnamefont {Carmelo}}, \bibinfo {editor} {\bibfnamefont {F.}~\bibnamefont {Guinea}},\ and\ \bibinfo {editor} {\bibfnamefont {E.}~\bibnamefont {Louis}}}\ (\bibinfo  {publisher} {Springer US},\ \bibinfo {address}
  {Boston, MA},\ \bibinfo {year} {1995})\ pp.\ \bibinfo {pages} {315--326}\BibitemShut {NoStop}%
\bibitem [{\citenamefont {Xiao}\ \emph {et~al.}(2022)\citenamefont {Xiao}, \citenamefont {Moreno}, \citenamefont {Fishman}, \citenamefont {Sels}, \citenamefont {Khatami},\ and\ \citenamefont {Scalettar}}]{Xiao2022}%
  \BibitemOpen
  \bibfield  {author} {\bibinfo {author} {\bibfnamefont {B.}~\bibnamefont {Xiao}}, \bibinfo {author} {\bibfnamefont {J.~R.}\ \bibnamefont {Moreno}}, \bibinfo {author} {\bibfnamefont {M.}~\bibnamefont {Fishman}}, \bibinfo {author} {\bibfnamefont {D.}~\bibnamefont {Sels}}, \bibinfo {author} {\bibfnamefont {E.}~\bibnamefont {Khatami}},\ and\ \bibinfo {author} {\bibfnamefont {R.}~\bibnamefont {Scalettar}},\ }\href@noop {} {\bibinfo {title} {{Extracting Off-Diagonal Order from Diagonal Basis Measurements}}} (\bibinfo {year} {2022}),\ \Eprint {https://arxiv.org/abs/2209.10565} {arXiv:2209.10565 [cond-mat.str-el]} \BibitemShut {NoStop}%
\bibitem [{\citenamefont {Zhang}\ and\ \citenamefont {Callaway}(1989)}]{zhang89}%
  \BibitemOpen
  \bibfield  {author} {\bibinfo {author} {\bibfnamefont {Y.}~\bibnamefont {Zhang}}\ and\ \bibinfo {author} {\bibfnamefont {J.}~\bibnamefont {Callaway}},\ }\href {https://doi.org/10.1103/PhysRevB.39.9397} {\bibfield  {journal} {\bibinfo  {journal} {Phys. Rev. B}\ }\textbf {\bibinfo {volume} {39}},\ \bibinfo {pages} {9397} (\bibinfo {year} {1989})}\BibitemShut {NoStop}%
\bibitem [{\citenamefont {Huang}\ \emph {et~al.}(2013)\citenamefont {Huang}, \citenamefont {Lai}, \citenamefont {Shi},\ and\ \citenamefont {Tsai}}]{Huang2013}%
  \BibitemOpen
  \bibfield  {author} {\bibinfo {author} {\bibfnamefont {W.-M.}\ \bibnamefont {Huang}}, \bibinfo {author} {\bibfnamefont {C.-Y.}\ \bibnamefont {Lai}}, \bibinfo {author} {\bibfnamefont {C.}~\bibnamefont {Shi}},\ and\ \bibinfo {author} {\bibfnamefont {S.-W.}\ \bibnamefont {Tsai}},\ }\href {https://doi.org/10.1103/PhysRevB.88.054504} {\bibfield  {journal} {\bibinfo  {journal} {Phys. Rev. B}\ }\textbf {\bibinfo {volume} {88}},\ \bibinfo {pages} {054504} (\bibinfo {year} {2013})}\BibitemShut {NoStop}%
\bibitem [{\citenamefont {Vandelli}\ \emph {et~al.}(2020)\citenamefont {Vandelli}, \citenamefont {Harkov}, \citenamefont {Stepanov}, \citenamefont {Gukelberger}, \citenamefont {Kozik}, \citenamefont {Rubio},\ and\ \citenamefont {Lichtenstein}}]{Vandelli20}%
  \BibitemOpen
  \bibfield  {author} {\bibinfo {author} {\bibfnamefont {M.}~\bibnamefont {Vandelli}}, \bibinfo {author} {\bibfnamefont {V.}~\bibnamefont {Harkov}}, \bibinfo {author} {\bibfnamefont {E.~A.}\ \bibnamefont {Stepanov}}, \bibinfo {author} {\bibfnamefont {J.}~\bibnamefont {Gukelberger}}, \bibinfo {author} {\bibfnamefont {E.}~\bibnamefont {Kozik}}, \bibinfo {author} {\bibfnamefont {A.}~\bibnamefont {Rubio}},\ and\ \bibinfo {author} {\bibfnamefont {A.~I.}\ \bibnamefont {Lichtenstein}},\ }\href {https://doi.org/10.1103/PhysRevB.102.195109} {\bibfield  {journal} {\bibinfo  {journal} {Phys. Rev. B}\ }\textbf {\bibinfo {volume} {102}},\ \bibinfo {pages} {195109} (\bibinfo {year} {2020})}\BibitemShut {NoStop}%
\bibitem [{\citenamefont {Wolf}\ \emph {et~al.}(2018)\citenamefont {Wolf}, \citenamefont {Schmidt},\ and\ \citenamefont {Rachel}}]{Wolf18}%
  \BibitemOpen
  \bibfield  {author} {\bibinfo {author} {\bibfnamefont {S.}~\bibnamefont {Wolf}}, \bibinfo {author} {\bibfnamefont {T.~L.}\ \bibnamefont {Schmidt}},\ and\ \bibinfo {author} {\bibfnamefont {S.}~\bibnamefont {Rachel}},\ }\href {https://doi.org/10.1103/PhysRevB.98.174515} {\bibfield  {journal} {\bibinfo  {journal} {Phys. Rev. B}\ }\textbf {\bibinfo {volume} {98}},\ \bibinfo {pages} {174515} (\bibinfo {year} {2018})}\BibitemShut {NoStop}%
\bibitem [{\citenamefont {Jiang}\ \emph {et~al.}(2018)\citenamefont {Jiang}, \citenamefont {H\"ahner}, \citenamefont {Schulthess},\ and\ \citenamefont {Maier}}]{Jiang2018}%
  \BibitemOpen
  \bibfield  {author} {\bibinfo {author} {\bibfnamefont {M.}~\bibnamefont {Jiang}}, \bibinfo {author} {\bibfnamefont {U.~R.}\ \bibnamefont {H\"ahner}}, \bibinfo {author} {\bibfnamefont {T.~C.}\ \bibnamefont {Schulthess}},\ and\ \bibinfo {author} {\bibfnamefont {T.~A.}\ \bibnamefont {Maier}},\ }\href {https://doi.org/10.1103/PhysRevB.97.184507} {\bibfield  {journal} {\bibinfo  {journal} {Phys. Rev. B}\ }\textbf {\bibinfo {volume} {97}},\ \bibinfo {pages} {184507} (\bibinfo {year} {2018})}\BibitemShut {NoStop}%
\bibitem [{\citenamefont {Kundu}\ and\ \citenamefont {Sénéchal}(2023)}]{kundu2023}%
  \BibitemOpen
  \bibfield  {author} {\bibinfo {author} {\bibfnamefont {S.}~\bibnamefont {Kundu}}\ and\ \bibinfo {author} {\bibfnamefont {D.}~\bibnamefont {Sénéchal}},\ }\href@noop {} {\bibinfo {title} {Cdmft+hfd : an extension of dynamical mean field theory for nonlocal interactions applied to the single band extended hubbard model}} (\bibinfo {year} {2023}),\ \Eprint {https://arxiv.org/abs/2310.16075} {arXiv:2310.16075 [cond-mat.str-el]} \BibitemShut {NoStop}%
\bibitem [{\citenamefont {Chen}\ \emph {et~al.}(2023)\citenamefont {Chen}, \citenamefont {Wang},\ and\ \citenamefont {Chen}}]{chen2022}%
  \BibitemOpen
  \bibfield  {author} {\bibinfo {author} {\bibfnamefont {W.-C.}\ \bibnamefont {Chen}}, \bibinfo {author} {\bibfnamefont {Y.}~\bibnamefont {Wang}},\ and\ \bibinfo {author} {\bibfnamefont {C.-C.}\ \bibnamefont {Chen}},\ }\href {https://doi.org/10.1103/PhysRevB.108.064514} {\bibfield  {journal} {\bibinfo  {journal} {Phys. Rev. B}\ }\textbf {\bibinfo {volume} {108}},\ \bibinfo {pages} {064514} (\bibinfo {year} {2023})}\BibitemShut {NoStop}%
\bibitem [{\citenamefont {Blankenbecler}\ \emph {et~al.}(1981)\citenamefont {Blankenbecler}, \citenamefont {Scalapino},\ and\ \citenamefont {Sugar}}]{Blankenbecler81}%
  \BibitemOpen
  \bibfield  {author} {\bibinfo {author} {\bibfnamefont {R.}~\bibnamefont {Blankenbecler}}, \bibinfo {author} {\bibfnamefont {D.~J.}\ \bibnamefont {Scalapino}},\ and\ \bibinfo {author} {\bibfnamefont {R.~L.}\ \bibnamefont {Sugar}},\ }\href {https://doi.org/10.1103/PhysRevD.24.2278} {\bibfield  {journal} {\bibinfo  {journal} {Phys. Rev. D}\ }\textbf {\bibinfo {volume} {24}},\ \bibinfo {pages} {2278} (\bibinfo {year} {1981})}\BibitemShut {NoStop}%
\bibitem [{\citenamefont {Hirsch}(1985)}]{hirsch85}%
  \BibitemOpen
  \bibfield  {author} {\bibinfo {author} {\bibfnamefont {J.~E.}\ \bibnamefont {Hirsch}},\ }\href {https://doi.org/10.1103/PhysRevB.31.4403} {\bibfield  {journal} {\bibinfo  {journal} {Phys. Rev. B}\ }\textbf {\bibinfo {volume} {31}},\ \bibinfo {pages} {4403} (\bibinfo {year} {1985})}\BibitemShut {NoStop}%
\bibitem [{\citenamefont {Scalettar}\ \emph {et~al.}(1989)\citenamefont {Scalettar}, \citenamefont {Bickers},\ and\ \citenamefont {Scalapino}}]{scalettar89}%
  \BibitemOpen
  \bibfield  {author} {\bibinfo {author} {\bibfnamefont {R.~T.}\ \bibnamefont {Scalettar}}, \bibinfo {author} {\bibfnamefont {N.~E.}\ \bibnamefont {Bickers}},\ and\ \bibinfo {author} {\bibfnamefont {D.~J.}\ \bibnamefont {Scalapino}},\ }\href {https://doi.org/10.1103/PhysRevB.40.197} {\bibfield  {journal} {\bibinfo  {journal} {Phys. Rev. B}\ }\textbf {\bibinfo {volume} {40}},\ \bibinfo {pages} {197} (\bibinfo {year} {1989})}\BibitemShut {NoStop}%
\bibitem [{\citenamefont {Kawashima}(2002)}]{Kawashima2002}%
  \BibitemOpen
  \bibfield  {author} {\bibinfo {author} {\bibfnamefont {N.}~\bibnamefont {Kawashima}},\ }\href {https://doi.org/10.1143/PTPS.145.138} {\bibfield  {journal} {\bibinfo  {journal} {Progress of Theoretical Physics Supplement}\ }\textbf {\bibinfo {volume} {145}},\ \bibinfo {pages} {138} (\bibinfo {year} {2002})}\BibitemShut {NoStop}%
\bibitem [{\citenamefont {dos Santos}(2003)}]{rrds2003}%
  \BibitemOpen
  \bibfield  {author} {\bibinfo {author} {\bibfnamefont {R.~R.}\ \bibnamefont {dos Santos}},\ }\href {https://doi.org/10.1590/S0103-97332003000100003} {\bibfield  {journal} {\bibinfo  {journal} {Braz. J. Phys}\ }\textbf {\bibinfo {volume} {33}},\ \bibinfo {pages} {63} (\bibinfo {year} {2003})}\BibitemShut {NoStop}%
\bibitem [{\citenamefont {Becca}\ and\ \citenamefont {Sorella}(2017)}]{sorella2017}%
  \BibitemOpen
  \bibfield  {author} {\bibinfo {author} {\bibfnamefont {F.}~\bibnamefont {Becca}}\ and\ \bibinfo {author} {\bibfnamefont {S.}~\bibnamefont {Sorella}},\ }\href {https://doi.org/10.1017/9781316417041} {\emph {\bibinfo {title} {{Quantum Monte Carlo Approaches for Correlated Systems}}}}\ (\bibinfo  {publisher} {Cambridge University Press},\ \bibinfo {year} {2017})\BibitemShut {NoStop}%
\bibitem [{\citenamefont {Sushchyev}\ and\ \citenamefont {Wessel}(2022)}]{Sushchyev2022}%
  \BibitemOpen
  \bibfield  {author} {\bibinfo {author} {\bibfnamefont {A.}~\bibnamefont {Sushchyev}}\ and\ \bibinfo {author} {\bibfnamefont {S.}~\bibnamefont {Wessel}},\ }\href {https://doi.org/10.1103/PhysRevB.106.155121} {\bibfield  {journal} {\bibinfo  {journal} {Phys. Rev. B}\ }\textbf {\bibinfo {volume} {106}},\ \bibinfo {pages} {155121} (\bibinfo {year} {2022})}\BibitemShut {NoStop}%
\bibitem [{\citenamefont {Yao}\ \emph {et~al.}(2022)\citenamefont {Yao}, \citenamefont {Wang},\ and\ \citenamefont {Wang}}]{Yao22}%
  \BibitemOpen
  \bibfield  {author} {\bibinfo {author} {\bibfnamefont {M.}~\bibnamefont {Yao}}, \bibinfo {author} {\bibfnamefont {D.}~\bibnamefont {Wang}},\ and\ \bibinfo {author} {\bibfnamefont {Q.-H.}\ \bibnamefont {Wang}},\ }\href {https://doi.org/10.1103/PhysRevB.106.195121} {\bibfield  {journal} {\bibinfo  {journal} {Phys. Rev. B}\ }\textbf {\bibinfo {volume} {106}},\ \bibinfo {pages} {195121} (\bibinfo {year} {2022})}\BibitemShut {NoStop}%
\bibitem [{\citenamefont {Mermin}\ and\ \citenamefont {Wagner}(1966)}]{Mermin1966}%
  \BibitemOpen
  \bibfield  {author} {\bibinfo {author} {\bibfnamefont {N.~D.}\ \bibnamefont {Mermin}}\ and\ \bibinfo {author} {\bibfnamefont {H.}~\bibnamefont {Wagner}},\ }\href {https://doi.org/10.1103/PhysRevLett.17.1133} {\bibfield  {journal} {\bibinfo  {journal} {Phys. Rev. Lett.}\ }\textbf {\bibinfo {volume} {17}},\ \bibinfo {pages} {1133} (\bibinfo {year} {1966})}\BibitemShut {NoStop}%
\bibitem [{\citenamefont {Hohenberg}(1967)}]{Hohenberg1967}%
  \BibitemOpen
  \bibfield  {author} {\bibinfo {author} {\bibfnamefont {P.~C.}\ \bibnamefont {Hohenberg}},\ }\href {https://doi.org/10.1103/PhysRev.158.383} {\bibfield  {journal} {\bibinfo  {journal} {Phys. Rev.}\ }\textbf {\bibinfo {volume} {158}},\ \bibinfo {pages} {383} (\bibinfo {year} {1967})}\BibitemShut {NoStop}%
\bibitem [{\citenamefont {Wessel}\ \emph {et~al.}(2017)\citenamefont {Wessel}, \citenamefont {Normand}, \citenamefont {Mila},\ and\ \citenamefont {Honecker}}]{Wessel2017}%
  \BibitemOpen
  \bibfield  {author} {\bibinfo {author} {\bibfnamefont {S.}~\bibnamefont {Wessel}}, \bibinfo {author} {\bibfnamefont {B.}~\bibnamefont {Normand}}, \bibinfo {author} {\bibfnamefont {F.}~\bibnamefont {Mila}},\ and\ \bibinfo {author} {\bibfnamefont {A.}~\bibnamefont {Honecker}},\ }\href {https://doi.org/10.21468/SciPostPhys.3.1.005} {\bibfield  {journal} {\bibinfo  {journal} {SciPost Phys.}\ }\textbf {\bibinfo {volume} {3}},\ \bibinfo {pages} {005} (\bibinfo {year} {2017})}\BibitemShut {NoStop}%
\bibitem [{\citenamefont {Mondaini}\ \emph {et~al.}(2022)\citenamefont {Mondaini}, \citenamefont {Tarat},\ and\ \citenamefont {Scalettar}}]{Mondaini2022}%
  \BibitemOpen
  \bibfield  {author} {\bibinfo {author} {\bibfnamefont {R.}~\bibnamefont {Mondaini}}, \bibinfo {author} {\bibfnamefont {S.}~\bibnamefont {Tarat}},\ and\ \bibinfo {author} {\bibfnamefont {R.~T.}\ \bibnamefont {Scalettar}},\ }\href {https://doi.org/10.1126/science.abg9299} {\bibfield  {journal} {\bibinfo  {journal} {Science}\ }\textbf {\bibinfo {volume} {375}},\ \bibinfo {pages} {418} (\bibinfo {year} {2022})}\BibitemShut {NoStop}%
\bibitem [{\citenamefont {Mondaini}\ \emph {et~al.}(2023)\citenamefont {Mondaini}, \citenamefont {Tarat},\ and\ \citenamefont {Scalettar}}]{Mondaini2023}%
  \BibitemOpen
  \bibfield  {author} {\bibinfo {author} {\bibfnamefont {R.}~\bibnamefont {Mondaini}}, \bibinfo {author} {\bibfnamefont {S.}~\bibnamefont {Tarat}},\ and\ \bibinfo {author} {\bibfnamefont {R.~T.}\ \bibnamefont {Scalettar}},\ }\href {https://doi.org/10.1103/PhysRevB.107.245144} {\bibfield  {journal} {\bibinfo  {journal} {Phys. Rev. B}\ }\textbf {\bibinfo {volume} {107}},\ \bibinfo {pages} {245144} (\bibinfo {year} {2023})}\BibitemShut {NoStop}%
\bibitem [{\citenamefont {Hirsch}(1983)}]{Hirsch83}%
  \BibitemOpen
  \bibfield  {author} {\bibinfo {author} {\bibfnamefont {J.~E.}\ \bibnamefont {Hirsch}},\ }\href {https://doi.org/10.1103/PhysRevB.28.4059} {\bibfield  {journal} {\bibinfo  {journal} {Phys. Rev. B}\ }\textbf {\bibinfo {volume} {28}},\ \bibinfo {pages} {4059} (\bibinfo {year} {1983})}\BibitemShut {NoStop}%
\bibitem [{\citenamefont {Golor}\ and\ \citenamefont {Wessel}(2015)}]{golor2015}%
  \BibitemOpen
  \bibfield  {author} {\bibinfo {author} {\bibfnamefont {M.}~\bibnamefont {Golor}}\ and\ \bibinfo {author} {\bibfnamefont {S.}~\bibnamefont {Wessel}},\ }\href {https://doi.org/10.1103/PhysRevB.92.195154} {\bibfield  {journal} {\bibinfo  {journal} {Phys. Rev. B}\ }\textbf {\bibinfo {volume} {92}},\ \bibinfo {pages} {195154} (\bibinfo {year} {2015})}\BibitemShut {NoStop}%
\bibitem [{\citenamefont {White}\ \emph {et~al.}(1989)\citenamefont {White}, \citenamefont {Scalapino}, \citenamefont {Sugar}, \citenamefont {Bickers},\ and\ \citenamefont {Scalettar}}]{white89}%
  \BibitemOpen
  \bibfield  {author} {\bibinfo {author} {\bibfnamefont {S.~R.}\ \bibnamefont {White}}, \bibinfo {author} {\bibfnamefont {D.~J.}\ \bibnamefont {Scalapino}}, \bibinfo {author} {\bibfnamefont {R.~L.}\ \bibnamefont {Sugar}}, \bibinfo {author} {\bibfnamefont {N.~E.}\ \bibnamefont {Bickers}},\ and\ \bibinfo {author} {\bibfnamefont {R.~T.}\ \bibnamefont {Scalettar}},\ }\href {https://doi.org/10.1103/PhysRevB.39.839} {\bibfield  {journal} {\bibinfo  {journal} {Phys. Rev. B}\ }\textbf {\bibinfo {volume} {39}},\ \bibinfo {pages} {839} (\bibinfo {year} {1989})}\BibitemShut {NoStop}%
\bibitem [{\citenamefont {Lima}\ \emph {et~al.}(2023)\citenamefont {Lima}, \citenamefont {Medeiros-Silva}, \citenamefont {dos Santos}, \citenamefont {Paiva},\ and\ \citenamefont {Costa}}]{Lima2023}%
  \BibitemOpen
  \bibfield  {author} {\bibinfo {author} {\bibfnamefont {L.~O.}\ \bibnamefont {Lima}}, \bibinfo {author} {\bibfnamefont {A.~R.}\ \bibnamefont {Medeiros-Silva}}, \bibinfo {author} {\bibfnamefont {R.~R.}\ \bibnamefont {dos Santos}}, \bibinfo {author} {\bibfnamefont {T.}~\bibnamefont {Paiva}},\ and\ \bibinfo {author} {\bibfnamefont {N.~C.}\ \bibnamefont {Costa}},\ }\href {https://doi.org/10.1103/PhysRevB.108.235163} {\bibfield  {journal} {\bibinfo  {journal} {Phys. Rev. B}\ }\textbf {\bibinfo {volume} {108}},\ \bibinfo {pages} {235163} (\bibinfo {year} {2023})}\BibitemShut {NoStop}%
\bibitem [{\citenamefont {Huse}(1988)}]{Huse88}%
  \BibitemOpen
  \bibfield  {author} {\bibinfo {author} {\bibfnamefont {D.~A.}\ \bibnamefont {Huse}},\ }\href {https://doi.org/10.1103/PhysRevB.37.2380} {\bibfield  {journal} {\bibinfo  {journal} {Phys. Rev. B}\ }\textbf {\bibinfo {volume} {37}},\ \bibinfo {pages} {2380} (\bibinfo {year} {1988})}\BibitemShut {NoStop}%
\bibitem [{\citenamefont {Nowadnick}\ \emph {et~al.}(2012)\citenamefont {Nowadnick}, \citenamefont {Johnston}, \citenamefont {Moritz}, \citenamefont {Scalettar},\ and\ \citenamefont {Devereaux}}]{Nowadnick12}%
  \BibitemOpen
  \bibfield  {author} {\bibinfo {author} {\bibfnamefont {E.~A.}\ \bibnamefont {Nowadnick}}, \bibinfo {author} {\bibfnamefont {S.}~\bibnamefont {Johnston}}, \bibinfo {author} {\bibfnamefont {B.}~\bibnamefont {Moritz}}, \bibinfo {author} {\bibfnamefont {R.~T.}\ \bibnamefont {Scalettar}},\ and\ \bibinfo {author} {\bibfnamefont {T.~P.}\ \bibnamefont {Devereaux}},\ }\href {https://doi.org/10.1103/PhysRevLett.109.246404} {\bibfield  {journal} {\bibinfo  {journal} {Phys. Rev. Lett.}\ }\textbf {\bibinfo {volume} {109}},\ \bibinfo {pages} {246404} (\bibinfo {year} {2012})}\BibitemShut {NoStop}%
\bibitem [{\citenamefont {Johnston}\ \emph {et~al.}(2013)\citenamefont {Johnston}, \citenamefont {Nowadnick}, \citenamefont {Kung}, \citenamefont {Moritz}, \citenamefont {Scalettar},\ and\ \citenamefont {Devereaux}}]{Johnston13}%
  \BibitemOpen
  \bibfield  {author} {\bibinfo {author} {\bibfnamefont {S.}~\bibnamefont {Johnston}}, \bibinfo {author} {\bibfnamefont {E.~A.}\ \bibnamefont {Nowadnick}}, \bibinfo {author} {\bibfnamefont {Y.~F.}\ \bibnamefont {Kung}}, \bibinfo {author} {\bibfnamefont {B.}~\bibnamefont {Moritz}}, \bibinfo {author} {\bibfnamefont {R.~T.}\ \bibnamefont {Scalettar}},\ and\ \bibinfo {author} {\bibfnamefont {T.~P.}\ \bibnamefont {Devereaux}},\ }\href {https://doi.org/10.1103/PhysRevB.87.235133} {\bibfield  {journal} {\bibinfo  {journal} {Phys. Rev. B}\ }\textbf {\bibinfo {volume} {87}},\ \bibinfo {pages} {235133} (\bibinfo {year} {2013})}\BibitemShut {NoStop}%
\bibitem [{\citenamefont {Costa}\ \emph {et~al.}(2020)\citenamefont {Costa}, \citenamefont {Seki}, \citenamefont {Yunoki},\ and\ \citenamefont {Sorella}}]{Costa20}%
  \BibitemOpen
  \bibfield  {author} {\bibinfo {author} {\bibfnamefont {N.}~\bibnamefont {Costa}}, \bibinfo {author} {\bibfnamefont {K.}~\bibnamefont {Seki}}, \bibinfo {author} {\bibfnamefont {S.}~\bibnamefont {Yunoki}},\ and\ \bibinfo {author} {\bibfnamefont {S.}~\bibnamefont {Sorella}},\ }\bibfield  {journal} {\bibinfo  {journal} {Communications Physics}\ }\textbf {\bibinfo {volume} {3}},\ \href {https://doi.org/10.1038/s42005-020-0342-2} {10.1038/s42005-020-0342-2} (\bibinfo {year} {2020})\BibitemShut {NoStop}%
\bibitem [{\citenamefont {Ferreira}\ \emph {et~al.}(2022)\citenamefont {Ferreira}, \citenamefont {Maciel}, \citenamefont {Vianna},\ and\ \citenamefont {Iemini}}]{Ferreira22}%
  \BibitemOpen
  \bibfield  {author} {\bibinfo {author} {\bibfnamefont {D.~L.~B.}\ \bibnamefont {Ferreira}}, \bibinfo {author} {\bibfnamefont {T.~O.}\ \bibnamefont {Maciel}}, \bibinfo {author} {\bibfnamefont {R.~O.}\ \bibnamefont {Vianna}},\ and\ \bibinfo {author} {\bibfnamefont {F.}~\bibnamefont {Iemini}},\ }\href {https://doi.org/10.1103/PhysRevB.105.115145} {\bibfield  {journal} {\bibinfo  {journal} {Phys. Rev. B}\ }\textbf {\bibinfo {volume} {105}},\ \bibinfo {pages} {115145} (\bibinfo {year} {2022})}\BibitemShut {NoStop}%
\bibitem [{\citenamefont {Xing}\ \emph {et~al.}(2021)\citenamefont {Xing}, \citenamefont {Chiu}, \citenamefont {Poletti}, \citenamefont {Scalettar},\ and\ \citenamefont {Batrouni}}]{Xing2021}%
  \BibitemOpen
  \bibfield  {author} {\bibinfo {author} {\bibfnamefont {B.}~\bibnamefont {Xing}}, \bibinfo {author} {\bibfnamefont {W.-T.}\ \bibnamefont {Chiu}}, \bibinfo {author} {\bibfnamefont {D.}~\bibnamefont {Poletti}}, \bibinfo {author} {\bibfnamefont {R.~T.}\ \bibnamefont {Scalettar}},\ and\ \bibinfo {author} {\bibfnamefont {G.}~\bibnamefont {Batrouni}},\ }\href {https://doi.org/10.1103/PhysRevLett.126.017601} {\bibfield  {journal} {\bibinfo  {journal} {Phys. Rev. Lett.}\ }\textbf {\bibinfo {volume} {126}},\ \bibinfo {pages} {017601} (\bibinfo {year} {2021})}\BibitemShut {NoStop}%
\bibitem [{\citenamefont {Dagotto}\ \emph {et~al.}(1994)\citenamefont {Dagotto}, \citenamefont {Riera}, \citenamefont {Chen}, \citenamefont {Moreo}, \citenamefont {Nazarenko}, \citenamefont {Alcaraz},\ and\ \citenamefont {Ortolani}}]{Dagotto94}%
  \BibitemOpen
  \bibfield  {author} {\bibinfo {author} {\bibfnamefont {E.}~\bibnamefont {Dagotto}}, \bibinfo {author} {\bibfnamefont {J.}~\bibnamefont {Riera}}, \bibinfo {author} {\bibfnamefont {Y.~C.}\ \bibnamefont {Chen}}, \bibinfo {author} {\bibfnamefont {A.}~\bibnamefont {Moreo}}, \bibinfo {author} {\bibfnamefont {A.}~\bibnamefont {Nazarenko}}, \bibinfo {author} {\bibfnamefont {F.}~\bibnamefont {Alcaraz}},\ and\ \bibinfo {author} {\bibfnamefont {F.}~\bibnamefont {Ortolani}},\ }\href {https://doi.org/10.1103/PhysRevB.49.3548} {\bibfield  {journal} {\bibinfo  {journal} {Phys. Rev. B}\ }\textbf {\bibinfo {volume} {49}},\ \bibinfo {pages} {3548} (\bibinfo {year} {1994})}\BibitemShut {NoStop}%
\bibitem [{\citenamefont {Micnas}\ \emph {et~al.}(1990)\citenamefont {Micnas}, \citenamefont {Ranninger},\ and\ \citenamefont {Robaszkiewicz}}]{Micnas90}%
  \BibitemOpen
  \bibfield  {author} {\bibinfo {author} {\bibfnamefont {R.}~\bibnamefont {Micnas}}, \bibinfo {author} {\bibfnamefont {J.}~\bibnamefont {Ranninger}},\ and\ \bibinfo {author} {\bibfnamefont {S.}~\bibnamefont {Robaszkiewicz}},\ }\href {https://doi.org/10.1103/RevModPhys.62.113} {\bibfield  {journal} {\bibinfo  {journal} {Rev. Mod. Phys.}\ }\textbf {\bibinfo {volume} {62}},\ \bibinfo {pages} {113} (\bibinfo {year} {1990})}\BibitemShut {NoStop}%
\bibitem [{\citenamefont {Fontenele}\ \emph {et~al.}(2022)\citenamefont {Fontenele}, \citenamefont {Costa}, \citenamefont {dos Santos},\ and\ \citenamefont {Paiva}}]{fontenele2022}%
  \BibitemOpen
  \bibfield  {author} {\bibinfo {author} {\bibfnamefont {R.~A.}\ \bibnamefont {Fontenele}}, \bibinfo {author} {\bibfnamefont {N.~C.}\ \bibnamefont {Costa}}, \bibinfo {author} {\bibfnamefont {R.~R.}\ \bibnamefont {dos Santos}},\ and\ \bibinfo {author} {\bibfnamefont {T.}~\bibnamefont {Paiva}},\ }\href {https://doi.org/10.1103/PhysRevB.105.184502} {\bibfield  {journal} {\bibinfo  {journal} {Phys. Rev. B}\ }\textbf {\bibinfo {volume} {105}},\ \bibinfo {pages} {184502} (\bibinfo {year} {2022})}\BibitemShut {NoStop}%
\bibitem [{\citenamefont {Fisher}(1971)}]{Fisher71}%
  \BibitemOpen
  \bibfield  {author} {\bibinfo {author} {\bibfnamefont {M.~E.}\ \bibnamefont {Fisher}},\ }in\ \href@noop {} {\emph {\bibinfo {booktitle} {{P}roceedings of the {E}nrico {F}ermi {I}nternational {S}chool of {P}hysics}}},\ Vol.~\bibinfo {volume} {51},\ \bibinfo {editor} {edited by\ \bibinfo {editor} {\bibfnamefont {M.~S.}\ \bibnamefont {Green}}}\ (\bibinfo  {publisher} {Academic Press, New York},\ \bibinfo {year} {1971})\BibitemShut {NoStop}%
\bibitem [{\citenamefont {Barber}(1983)}]{Barber83}%
  \BibitemOpen
  \bibfield  {author} {\bibinfo {author} {\bibfnamefont {M.~N.}\ \bibnamefont {Barber}},\ }in\ \href@noop {} {\emph {\bibinfo {booktitle} {Phase Transitions and Critical Phenomena}}},\ Vol.~\bibinfo {volume} {8},\ \bibinfo {editor} {edited by\ \bibinfo {editor} {\bibfnamefont {C.}~\bibnamefont {Domb}}\ and\ \bibinfo {editor} {\bibfnamefont {J.~L.}\ \bibnamefont {Lebowitz}}}\ (\bibinfo  {publisher} {Academic Press},\ \bibinfo {address} {New York},\ \bibinfo {year} {1983})\ p.\ \bibinfo {pages} {145}\BibitemShut {NoStop}%
\bibitem [{\citenamefont {dos Santos}\ and\ \citenamefont {Sneddon}(1981)}]{dosSantos81a}%
  \BibitemOpen
  \bibfield  {author} {\bibinfo {author} {\bibfnamefont {R.~R.}\ \bibnamefont {dos Santos}}\ and\ \bibinfo {author} {\bibfnamefont {L.}~\bibnamefont {Sneddon}},\ }\href {https://doi.org/10.1103/PhysRevB.23.3541} {\bibfield  {journal} {\bibinfo  {journal} {Phys. Rev. B}\ }\textbf {\bibinfo {volume} {23}},\ \bibinfo {pages} {3541} (\bibinfo {year} {1981})}\BibitemShut {NoStop}%
\bibitem [{\citenamefont {Stanley}(1971)}]{Stanley71}%
  \BibitemOpen
  \bibfield  {author} {\bibinfo {author} {\bibfnamefont {H.~E.}\ \bibnamefont {Stanley}},\ }\href@noop {} {\emph {\bibinfo {title} {Introduction to Phase Transitions and Critical Phenomena}}},\ International series of monographs on physics\ (\bibinfo  {publisher} {Oxford University Press},\ \bibinfo {year} {1971})\BibitemShut {NoStop}%
\bibitem [{\citenamefont {\ifmmode~\check{S}\else \v{S}\fi{}untajs}\ \emph {et~al.}(2020)\citenamefont {\ifmmode~\check{S}\else \v{S}\fi{}untajs}, \citenamefont {Bon\ifmmode~\check{c}\else \v{c}\fi{}a}, \citenamefont {Prosen},\ and\ \citenamefont {Vidmar}}]{Suntajs2020}%
  \BibitemOpen
  \bibfield  {author} {\bibinfo {author} {\bibfnamefont {J.}~\bibnamefont {\ifmmode~\check{S}\else \v{S}\fi{}untajs}}, \bibinfo {author} {\bibfnamefont {J.}~\bibnamefont {Bon\ifmmode~\check{c}\else \v{c}\fi{}a}}, \bibinfo {author} {\bibfnamefont {T.~c.~v.}\ \bibnamefont {Prosen}},\ and\ \bibinfo {author} {\bibfnamefont {L.}~\bibnamefont {Vidmar}},\ }\href {https://doi.org/10.1103/PhysRevB.102.064207} {\bibfield  {journal} {\bibinfo  {journal} {Phys. Rev. B}\ }\textbf {\bibinfo {volume} {102}},\ \bibinfo {pages} {064207} (\bibinfo {year} {2020})}\BibitemShut {NoStop}%
\bibitem [{\citenamefont {Feng}\ and\ \citenamefont {Scalettar}(2020)}]{feng2020}%
  \BibitemOpen
  \bibfield  {author} {\bibinfo {author} {\bibfnamefont {C.}~\bibnamefont {Feng}}\ and\ \bibinfo {author} {\bibfnamefont {R.~T.}\ \bibnamefont {Scalettar}},\ }\href {https://doi.org/10.1103/PhysRevB.102.235152} {\bibfield  {journal} {\bibinfo  {journal} {Phys. Rev. B}\ }\textbf {\bibinfo {volume} {102}},\ \bibinfo {pages} {235152} (\bibinfo {year} {2020})}\BibitemShut {NoStop}%
\bibitem [{\citenamefont {Nelson}\ and\ \citenamefont {Kosterlitz}(1977)}]{Nelson77}%
  \BibitemOpen
  \bibfield  {author} {\bibinfo {author} {\bibfnamefont {D.~R.}\ \bibnamefont {Nelson}}\ and\ \bibinfo {author} {\bibfnamefont {J.~M.}\ \bibnamefont {Kosterlitz}},\ }\href {https://doi.org/10.1103/PhysRevLett.39.1201} {\bibfield  {journal} {\bibinfo  {journal} {Phys. Rev. Lett.}\ }\textbf {\bibinfo {volume} {39}},\ \bibinfo {pages} {1201} (\bibinfo {year} {1977})}\BibitemShut {NoStop}%
\bibitem [{\citenamefont {Scalapino}\ \emph {et~al.}(1992)\citenamefont {Scalapino}, \citenamefont {White},\ and\ \citenamefont {Zhang}}]{Scalapino92}%
  \BibitemOpen
  \bibfield  {author} {\bibinfo {author} {\bibfnamefont {D.~J.}\ \bibnamefont {Scalapino}}, \bibinfo {author} {\bibfnamefont {S.~R.}\ \bibnamefont {White}},\ and\ \bibinfo {author} {\bibfnamefont {S.~C.}\ \bibnamefont {Zhang}},\ }\href {https://doi.org/10.1103/PhysRevLett.68.2830} {\bibfield  {journal} {\bibinfo  {journal} {Phys. Rev. Lett.}\ }\textbf {\bibinfo {volume} {68}},\ \bibinfo {pages} {2830} (\bibinfo {year} {1992})}\BibitemShut {NoStop}%
\bibitem [{\citenamefont {Scalapino}\ \emph {et~al.}(1993)\citenamefont {Scalapino}, \citenamefont {White},\ and\ \citenamefont {Zhang}}]{Scalapino93}%
  \BibitemOpen
  \bibfield  {author} {\bibinfo {author} {\bibfnamefont {D.~J.}\ \bibnamefont {Scalapino}}, \bibinfo {author} {\bibfnamefont {S.~R.}\ \bibnamefont {White}},\ and\ \bibinfo {author} {\bibfnamefont {S.}~\bibnamefont {Zhang}},\ }\href {https://doi.org/10.1103/PhysRevB.47.7995} {\bibfield  {journal} {\bibinfo  {journal} {Phys. Rev. B}\ }\textbf {\bibinfo {volume} {47}},\ \bibinfo {pages} {7995} (\bibinfo {year} {1993})}\BibitemShut {NoStop}%
\bibitem [{\citenamefont {Paiva}\ \emph {et~al.}(2004)\citenamefont {Paiva}, \citenamefont {dos Santos}, \citenamefont {Scalettar},\ and\ \citenamefont {Denteneer}}]{Paiva04}%
  \BibitemOpen
  \bibfield  {author} {\bibinfo {author} {\bibfnamefont {T.}~\bibnamefont {Paiva}}, \bibinfo {author} {\bibfnamefont {R.~R.}\ \bibnamefont {dos Santos}}, \bibinfo {author} {\bibfnamefont {R.~T.}\ \bibnamefont {Scalettar}},\ and\ \bibinfo {author} {\bibfnamefont {P.~J.~H.}\ \bibnamefont {Denteneer}},\ }\href {https://doi.org/10.1103/PhysRevB.69.184501} {\bibfield  {journal} {\bibinfo  {journal} {Phys. Rev. B}\ }\textbf {\bibinfo {volume} {69}},\ \bibinfo {pages} {184501} (\bibinfo {year} {2004})}\BibitemShut {NoStop}%
\bibitem [{\citenamefont {dos Santos}(1993)}]{dosSantos93}%
  \BibitemOpen
  \bibfield  {author} {\bibinfo {author} {\bibfnamefont {R.~R.}\ \bibnamefont {dos Santos}},\ }\href {https://doi.org/10.1103/PhysRevB.48.3976} {\bibfield  {journal} {\bibinfo  {journal} {Phys. Rev. B}\ }\textbf {\bibinfo {volume} {48}},\ \bibinfo {pages} {3976} (\bibinfo {year} {1993})}\BibitemShut {NoStop}%
\bibitem [{\citenamefont {Sun}\ and\ \citenamefont {Lin}(2024)}]{Sun24}%
  \BibitemOpen
  \bibfield  {author} {\bibinfo {author} {\bibfnamefont {Z.}~\bibnamefont {Sun}}\ and\ \bibinfo {author} {\bibfnamefont {H.-Q.}\ \bibnamefont {Lin}},\ }\href {https://doi.org/10.1103/PhysRevB.109.035107} {\bibfield  {journal} {\bibinfo  {journal} {Phys. Rev. B}\ }\textbf {\bibinfo {volume} {109}},\ \bibinfo {pages} {035107} (\bibinfo {year} {2024})}\BibitemShut {NoStop}%
\bibitem [{\citenamefont {Mattheiss}(1987)}]{Mattheiss87}%
  \BibitemOpen
  \bibfield  {author} {\bibinfo {author} {\bibfnamefont {L.~F.}\ \bibnamefont {Mattheiss}},\ }\href {https://doi.org/10.1103/PhysRevLett.58.1028} {\bibfield  {journal} {\bibinfo  {journal} {Phys. Rev. Lett.}\ }\textbf {\bibinfo {volume} {58}},\ \bibinfo {pages} {1028} (\bibinfo {year} {1987})}\BibitemShut {NoStop}%
\bibitem [{\citenamefont {Rademaker}\ \emph {et~al.}(2013)\citenamefont {Rademaker}, \citenamefont {Johnston}, \citenamefont {Zaanen},\ and\ \citenamefont {van~den Brink}}]{Rademaker2013}%
  \BibitemOpen
  \bibfield  {author} {\bibinfo {author} {\bibfnamefont {L.}~\bibnamefont {Rademaker}}, \bibinfo {author} {\bibfnamefont {S.}~\bibnamefont {Johnston}}, \bibinfo {author} {\bibfnamefont {J.}~\bibnamefont {Zaanen}},\ and\ \bibinfo {author} {\bibfnamefont {J.}~\bibnamefont {van~den Brink}},\ }\href {https://doi.org/10.1103/PhysRevB.88.235115} {\bibfield  {journal} {\bibinfo  {journal} {Phys. Rev. B}\ }\textbf {\bibinfo {volume} {88}},\ \bibinfo {pages} {235115} (\bibinfo {year} {2013})}\BibitemShut {NoStop}%
\bibitem [{\citenamefont {Gubernatis}\ \emph {et~al.}(2016)\citenamefont {Gubernatis}, \citenamefont {Kawashima},\ and\ \citenamefont {Werner}}]{gubernatis2016}%
  \BibitemOpen
  \bibfield  {author} {\bibinfo {author} {\bibfnamefont {J.}~\bibnamefont {Gubernatis}}, \bibinfo {author} {\bibfnamefont {N.}~\bibnamefont {Kawashima}},\ and\ \bibinfo {author} {\bibfnamefont {P.}~\bibnamefont {Werner}},\ }\bibinfo {title} {Determinant method},\ in\ \href {https://doi.org/10.1017/CBO9780511902581.008} {\emph {\bibinfo {booktitle} {Quantum Monte Carlo Methods: Algorithms for Lattice Models}}}\ (\bibinfo  {publisher} {Cambridge University Press},\ \bibinfo {year} {2016})\ p.\ \bibinfo {pages} {180–213}\BibitemShut {NoStop}%
\end{thebibliography}%

\end{document}